\documentstyle[aps,epsfig]{revtex}
\begin{document}
\title{Self force on a scalar charge in the spacetime of a stationary,
axisymmetric black hole}
\author{Lior M. Burko and Yuk Tung Liu}
\address{Theoretical Astrophysics, California Institute of
Technology, Pasadena, California 91125}
\date{\today}
\maketitle
%\draft

\begin{abstract}
We study the self force acting on a particle 
endowed with scalar charge, which is held static (with 
respect to an undragged, static observer at infinity) outside a stationary, 
axially-symmetric black hole. We find that the acceleration due to the 
self force is in the same direction as the black hole's spin, and diverges when 
the particle approaches the outer boundary of the black hole's ergosphere. 
This acceleration diverges more rapidly approaching the ergosphere's 
boundary than the particle's acceleration in the absence of the self force. 
At the leading order this self force is a (post)$^2$-Newtonian effect.
For scalar charges with high charge-to-mass ratio, the acceleration due to the self 
force starts dominating over the regular acceleration already far from the 
black hole. The self force is proportional to the rate at which the black hole's 
rotational energy is dissipated. This self force is local 
(i.e., only the Abraham-Lorentz-Dirac force and the local coupling to 
Ricci curvature contribute to it). The non-local, tail part of the self force is zero.
\newline
\newline
PACS number(s): 04.25.-g, 04.70.-s, 04.70.Bw 
\end{abstract}

\section{Introduction and Summary}

Recently there has been much interest in the calculation of self
interaction of particles in curved spacetime. This growing interest is
motivated by the prospects of detection of low-frequency gravitational
waves in the not-so-distant future by space-borne gravitational wave
detectors such as LISA. The main challenge is to compute the orbital
evolution, and the resulting wave forms, from a compact object orbiting a
supermassive central black hole in the extreme mass ratio case. The motion
of the compact object, which may be construed as a structureless particle,
is geodesic in the limit of zero mass. However, when the particle is
endowed  with a finite (albeit small) mass, its motion is changed. 
Spacetime is now determined by the energy-momenta of both the black
hole and the particle, and the latter follows then a geodesic of 
the new spacetime, which is perturbed by its own energy-momentum relative 
to the original spacetime. An alternative viewpoint is to consider the
motion of the particle as an accelerated, non-geodesic motion in the
unperturbed spacetime of the central black hole. Whereas this latter
approach is less in the spirit of Einstein's General Relativity, which
``eliminated'' gravitational forces in favor of geometry, than the former
approach, it has the advantage that the unperturbed spacetime is often
very simple. (For many interesting cases it is, e.g., the 
stationary and axially-symmetric spacetime of a Kerr black hole.) We are
thus led naturally to translate the problem of finding the orbital
evolution of the particle to the following question: what are all the
momentary forces which act on the compact object? (In the absence of
external forces, there would be just the self force.) Obviously, knowledge of
all the forces which act on an object in a given spacetime allows for the
computation of the orbit and consequently also for the computation of the
emitted gravitational waves. 

The problem of finding the self force which acts on a particle in curved
spacetime is not easy. The reason is that in curved spacetime, due
to the failure of the Huygens principle, the retarded Green's function
associated with the particle's field has support inside the future
light cone, and in particular also on the future world line. (The physical
origin for this phenomenon is the scattering of the emitted waves off the
spacetime curvature.) The part of the Green's function inside the light
cone has been dubbed its ``tail'' part, and it is the calculation of this
tail part of the retarded Green's function which is the greatest challenge
in the computation of the self force (a.k.a.\ the radiation reaction
force). A difficulty associated with the calculation of the tail part of
the Green's function is the prescription used in order to separate the
tail part, which affects the motion of the particle, from the
instantaneous, divergent part, which arises from the typical divergence
of the particle's field in the coincidence limit of the source for the
field and the evaluation point. 

Several approaches have been proposed for the regularization of the self
force (SF). One approach was first suggested by Dirac for a pointlike
electric charge in flat spacetime \cite{dirac-38}, and later used by
DeWitt and Brehme for the case of an electric charge in curved spacetime
\cite{dewitt-brehme-60} and by Mino, Sasaki, and Tanaka for the case of a
pointlike particle coupled to linearized gravity \cite{mino-etal-97}. 
The idea is
to impose local energy and momentum conservation on a tube surrounding the
particle's world line, and to integrate the conservation laws across the
tube, thus obtaining the particle's equations of motion, including the SF 
effects. The divergent piece of the self force is then
removed by a mass-regularization procedure. A second, axiomatic approach,
which leads to the same expression as the approach described above, was
proposed by Quinn and Wald for electromagnetic and 
gravitational self forces \cite{quinn-wald-97} and by Quinn for the
scalar field case \cite{quinn-00}.
According to this approach, the regularization of the SF is
performed by comparing the forces in two different spacetimes. 

For the case of a particle coupled to a (minimally-coupled, massless)
scalar field, the total SF which acts on the particle is given by
\cite{quinn-00}
\begin{equation}
F_{\mu}^{\rm SF}=\frac{1}{3}q^2\left(\ddot{u}_{\mu}
-u_{\mu}\dot{u}_{\alpha}\dot{u}^{\alpha}\right)
+\frac{1}{6}q^2{\cal R}_{\mu}+\lim_{\epsilon\to
0^-}q^2\int_{-\infty}^{\tau+\epsilon}\,d\tau'
\nabla_{\mu}G^{\rm ret}
[z^{\alpha}(\tau),z^{\alpha}(\tau')]\, ,
\label{formal-expression}
\end{equation} 
where ${\cal
R}_{\mu}=R_{\mu\alpha}u^{\alpha}+R_{\alpha\beta}u^{\alpha}u^{\beta}u_{\mu}
-\frac{1}{2}Ru_{\mu}$. Here, $R_{\mu\nu}$ is the Ricci tensor, $R$ is the
curvature scalar, and $u^{\alpha}$ is the particle's four-velocity. An
overdot denotes (covariant) differentiation with respect to proper time
$\tau$, $q$ is the particle's scalar charge, $G^{\rm 
ret}[z^{\alpha}(\tau),z^{\alpha}(\tau')]$ is the retarded Green's function, and 
$z^{\alpha}(\tau)$ is the
particle's world line. The total force which acts on the particle is the
sum of external forces (e.g., forces which result from external scalar
fields) and the SF (\ref{formal-expression}). 
The SF in Eq.\ (\ref{formal-expression}) 
has three contributions: the first is a local, Abraham-Lorentz-Dirac (ALD)
type force. The ALD force consists of two terms, a term proportional to
the proper time derivative of the four-acceleration, called the ``Schott
part'' of the ALD force, and a term proportional to the four-acceleration
squared, which we shall call here the ``damping part.'' The second term of
Eq.\ (\ref{formal-expression}) comes from local coupling of the particle
to Ricci curvature. This term preserves the conformal invariance of the
SF. The two local terms, namely the ALD and Ricci-coupled terms,
constitute the local part of the SF, $F_{\mu}^{\rm local}$. 
The third term in the SF is the non-local tail term, which involves
an integration
over the gradient of the retarded Green's function along the entire past
world line of the particle. 

A number of practical methods have been used in order to find the effects
of the tail part of the SF. One such approach avoids the calculation of
the SF by applying balance arguments to quantities which are constants of
motion in the absence of radiation reaction (i.e., the energy and angular 
momentum). This method has been successful to quite high relativistic
order when the black hole is
spherically symmetric, or when the motion is circular or equatorial around
a spinning black hole, because in these cases the orbital evolution is
uniquely determined by the rate of change of energy and angular
momentum, and global conservation laws imply that the fluxes of energy and
angular momentum across a distant sphere and down the event horizon of the
black hole equal the rate of loss of energy and angular momentum by the
particle, respectively.  However, this method is useless for cases where
there is
non-trivial evolution of Carter's constant, which is a non-additive
constant of motion in the absence of radiation reaction. This method has
in addition two other problems. First, it involves averaging over the
orbital period. Thus, when the orbital evolution is fast and the time
scale for the orbital evolution becomes comparable with the time scale for
the orbital period, this method becomes inaccurate. Second, this method is
sensitive to the slowest decaying part of the field at large distances.
Consequently, it handles well the dissipative, radiative part of the
field, but completely ignores the conservative part of the SF. Although
the latter does not cause any net loss of energy and angular momentum, it
is still important for the orbital evolution \cite{burko-ijmpa}. In fact,
even in the absence of dissipation, the conservative SF pushes the
particle off the geodesic, and thus causes orbital evolution which may be
of practical importance. 

A second approach is based on the radiative Green's function
\cite{dirac-38}. Specifically, one can write the
retarded Green's function $G^{\rm ret}$ as the sum of two terms, namely
$$G^{\rm ret}=\frac{1}{2}\left(G^{\rm ret}+G^{\rm adv}\right)
+\frac{1}{2}\left(G^{\rm ret}-G^{\rm adv}\right),$$
where $G^{\rm adv}$ is the advanced Green's function. The first term is
time symmetric, and consequently does not include the radiative part of 
the field. Instead, it relates to the non-radiative Coulomb piece of the
field, which is the source for the divergence. If one considers then only the
second term, namely the radiative Green's function, one can obtain an 
expression for the field of the charge, which is finite on the world line
of the charge, and from this field obtain the SF. Although
this approach is very successful in flat spacetime, it suffers from an
inherent difficulty in curved spacetime. Specifically, in curved spacetime
it is anti-causal: The radiative Green's function includes the advanced
Green's function, which in curved spacetime has support inside the future
light cone. Consequently, the momentary force on the particle depends in
principle, according to the radiative Green's function, on the
entire future history of the particle. When the future history is
completely known, e.g., an eternal static particle, or eternal circular
motion, this approach is expected to yield correct results. However, in
general the future history is unknown, and might also be subjected to free
will, such that an approach built on the radiative Green's function is
unsatisfactory. An approach which is based solely on the causal retarded
field is clearly preferable. (In what follows we shall indeed base our
calculations on the retarded field.) Also this method ignores the
conservative piece of the SF, which is included in the discarded
time-symmetric part of $G^{\rm ret}$.

This approach was used by Gal'tsov to obtain the SF
on scalar, electric, and gravitational charges in the spacetime of a Kerr
black hole \cite{galtsov-82}. In particular, Gal'tsov found the SF acting
on charges in a uniform circular orbit around a Kerr black hole in the weak
field limit, and studied also the forces on static charges in the same
limit. 

In this Paper we focus on the SF acting on a scalar charge. Although a
scalar charge is just a toy model for the more interesting and more
realistic gravitational charge (a point mass), it already involves much of
the properties of more realistic fields, especially the tail part of the
SF. Yet, some of the complications associated with the gravitational SF
are not invoked, notably the gauge problem of the SF. Also, the scalar
field Green's function has just one component, which simplifies the
analysis. Thus, despite its relative simplicity, the problem of the scalar
field SF captures the essence of the physics, while avoiding some
technical complications. For that reason the scalar field SF has been a
very useful toy model.

In what follows we shall compute the self
force to linear order in the particle's field [i.e., to order
(charge)$^2$, and neglect corrections of order (charge)$^4$ or higher.
We thus assume that the particle's charge is much smaller than the typical
length scale for the gravitational field, specifically the black hole's
mass.]                                         

For a number of simple cases, e.g., static electric \cite{linet-76} or
scalar \cite{wiseman-00} charges in
the spacetime of a Schwarzschild black hole, or a static electric charge
on the polar axis of a Kerr black hole \cite{leaute-linet-82}, the
particle's field is known exactly in a closed form. Indeed, these exact
solutions were used to find the SF for those cases
\cite{frolov-zelnikov-82,wiseman-00,leaute-linet-82}. However, in most
cases an exact solution for the particle's field is unknown. For
sufficiently simple spacetimes, e.g., that of a stationary and
axisymmetric black hole, one can decompose the field into Fourier-harmonic
modes, which can be obtained relatively easily. Specifically, the
individual modes of
the field (or the Green's function) satisfy an ordinary differential
equation (whereas the field itself satisfies a partial differential
equation), which for almost all cases can be solved (at least numerically 
using standard methods). 
In addition, it turns out that the individual modes of the field are
continuous across the particle's world line and the resulting
contributions to the SF of the individual modes are bounded.
The divergency arises only at the step of summation
over all modes. 

This prompted Ori to propose a calculation of the SF
effects which is based on the retarded field and on a mode decomposition
\cite{ori-95} (in that case
the adiabatic, orbit integrated, evolution rate of the constants of
motion in Kerr). More recently, Ori proposed to apply that method directly
for the calculation of the SF \cite{ori-unpublished}. The greatest
challenge, as was already mentioned, lies with an appropriate prescription
for the regularization of the mode sum. When the  
individual modes of the SF are summed na\"{\i}vely, the result 
typically diverges. The reason for this
divergence is that the modes do not distinguish between the tail and the
instantaneous parts of the SF, and contribute to both. [This occurs
already in the case of a static scalar or electric charge in flat
spacetime, when the position of the charge does not coincide with the
center of the coordinates. In that case the contribution to the SF of each
mode (after summation over all azimuthal numbers $m$ from $-l$ to $l$) is
independent of the mode number $l$, and is given by $-q^2/(2r^2)$, $q$
being the (scalar or electric) charge, and $r$ being the position of the
charge. Obviously, the sum over modes diverges, and should be removed by a
certain regularization prescription.] A mode sum regularization
prescription (MSRP), which handles this divergence, and which is an
application of the approaches of Mino {\it et al} \cite{mino-etal-97} and
of Quinn and Wald \cite{quinn-wald-97}, was proposed by Ori
\cite{ori-unpublished,barack-ori}. 

Next, we describe the MSRP very succinctly. Further details are presented 
in Refs.\ \cite{barack-ori,barack-00}. The contribution to the physical SF 
from the tail part of the Green's function can be decomposed into
stationary Teukolsky modes, and then summed over the frequencies $\omega$
and the azimuthal numbers $m$. The tail part of the SF equals then the limit
$\epsilon\to 0^-$ of the sum over all $l$ modes, 
of the difference between the force
sourced by the entire world line (the bare force ${^{\rm 
bare}}f_{\mu}^{l}$) and
the force sourced by the half-infinite world line
to the future of $\epsilon$, where the particle has proper time
$\tau=0$, and $\tau=\epsilon$ is an event along the past ($\tau<0$)
world line. Next, we seek a regularization function $h^{l}_{\mu}$ which is
independent of $\epsilon$, such that the series $\sum_{l}({^{\rm
bare}}f_{\mu}^{l}-h^{l}_{\mu})$ converges. Once such a function is found,
the regularized tail part of the SF is then given by 
\begin{equation}
{^{\rm tail}}F_{\mu}\equiv\lim_{\epsilon\to
0^-}q^2\int_{-\infty}^{\tau+\epsilon}\,d\tau'
\nabla_{\mu}G^{\rm ret}
[z^{\alpha}(\tau),z^{\alpha}(\tau')]=\sum_{l=0}^{\infty}\left({^{\rm
bare}}f_{\mu}^{l}-h^{l}_{\mu}\right)-d_{\mu}\, ,
\end{equation} 
where $d_{\mu}$ is a finite
valued function. MSRP then shows, from a local integration of the Green's
function, that the regularization function
$h^{l}_{\mu}=a_{\mu}l+b_{\mu}+c_{\mu}l^{-1}$.  For several cases, which
have already been studied, MSRP yields the values of the functions
$a_{\mu}, b_{\mu}, c_{\mu}$ and $d_{\mu}$ analytically. Alternatively,
$a_{\mu}, b_{\mu}$, and $c_{\mu}$ (but not $d_{\mu}$) can also be found
from the large-$l$ behavior of ${^{\rm bare}}f_{\mu}^{l}$. 
As $\sum_{l=0}^{\infty}\left({^{\rm bare}}f_{\mu}^{l}-h^{l}_{\mu}\right)$
converges, it is clear that the large-$l$ behavior of ${^{\rm
bare}}f_{\mu}^{l}$ is identical to the large-$l$ behavior of $h^l_{\mu}$.  
For more
details on MSRP, and in particular on the local integration of the Green's
function and the analytical derivation of the MSRP parameters, see Refs.\ 
\cite{barack-ori,barack-00}.        

The MSRP has been applied successfully for a number of cases, including
the SF on static scalar or electric charges in the spacetime of a
Schwarzschild black hole \cite{burko-cqg}, a scalar charge in uniform
circular orbit around a Schwarzschild black hole \cite{burko-prl}, an
electric charge in uniform circular motion in Minkowski spacetime
\cite{burko-ajp}, a scalar charge which is in radial free fall in the
absence of 
the SF into a Schwarzschild black hole \cite{barack-burko-00}, and the SF
on static scalar or electric charges inside or outside thin spherical
shells  \cite{shells-00}. In all these cases the MSRP parameters 
$a_{\mu}, b_{\mu}, c_{\mu}$ and $d_{\mu}$ were known analytically, and
were used in the regularization of the SF (for a list of these parameters
see Ref.\ \cite{burko-amaldi}). The SF has not been calculated using
MSRP yet for
cases where the MSRP parameters are unknown. However, there are
interesting cases where these parameters have not been found analytically
yet. Specifically, they have not been found analytically for any spacetime
which is not spherically symmetric. In addition, when the spacetime of a
spinning black hole is concerned, their analytical derivation is
expected to be considerably more complicated than in the spherically
symmetric case: the Green's function is generally 
time dependent, and for the corresponding wave evolution in the spacetime 
of a rotating black hole different $l$ modes of the field couple \cite{barack-99}, 
a phenomenon which does not happen in spherically-symmetric spacetimes. As 
was noted
above, the MSRP parameters $a_{\mu}, b_{\mu}$, and $c_{\mu}$ can be found
also by studying the large-$l$ behavior of the individual modes of the
bare force.
However, without knowledge of the MSRP parameter $d_{\mu}$ any regularized
result will not be unambiguous. It appears, then, that a local analysis of
the retarded Green's function is unavoidable. Remarkably, it has been
found that in all the cases for which the MSRP parameter $d_{\mu}$ is
known, it satisfies a simple relation with the local part of the SF.
Specifically, it has been shown that in those cases $d_{\mu}$ equals the
sum of the ALD force and the Ricci-curvature coupled piece of the SF,
i.e., in the case of a scalar charge,  
\begin{equation}
d_{\mu}=\frac{1}{3}q^2\left(\ddot{u}_{\mu} 
-u_{\mu}\dot{u}_{\alpha}\dot{u}^{\alpha}\right)
+\frac{1}{6}q^2{\cal R}_{\mu} \, .
\label{d-conjecture}
\end{equation}
It was then conjectured, based on the particular cases for which it was
found to be true, that Eq.\ (\ref{d-conjecture}) is generally satisfied,
at least for large classes of scalar charges
\cite{barack-00,burko-amaldi}. (In Ref.\ \cite{barack-00} this conjecture 
applies only to static spherically-symmetric spacetimes. Here we expand 
the domain of validity of this conjecture to include at the least 
also the spacetime of a stationary, axially-symmetric black hole. We 
note that if this conjecture is found to be valid in general, it would not 
be unexpected.) If this conjecture is true in general, the
full SF is given by just 
\begin{equation}
F_{\mu}=\sum_{l=0}^{\infty}\left({^{\rm  
bare}}f_{\mu}^{l}-h^{l}_{\mu}\right)\, ,
\label{full-SF}
\end{equation}
and all the terms appearing in Eq.\ (\ref{full-SF}) can be found by
studying the individual modes of the {\it bare} SF only. [Even if this
conjecture is not true in general, Eq.\ (\ref{full-SF}) still holds for the
classes of cases for which the conjecture is true.] 

In this Paper we shall assume that Eq.\ (\ref{d-conjecture}) holds for the
case of a static scalar charge in the spacetime of a stationary,
axisymmetric black hole, and compute the SF for that case. By a static
particle we mean a particle whose Boyer-Lindquist coordinates
$r,\theta$, and $\varphi$ are fixed. Such a particle is static with
respect to an undragged, static observer at
infinity, but it rotates in the opposite direction to the black hole's
spin with respect to a freely falling local observer. This is the
first application of MSRP for a spacetime which is not spherically
symmetric. In addition, we also use MSRP without prior knowledge of the 
regularization parameters. 
By comparing our result for the SF with known results for a
Kerr spacetime in the weak field limit (which were obtained using
independent methods in Ref.\ \cite{galtsov-82}), we shall, in fact, prove
the validity of Eq.\ (\ref{d-conjecture}) in that limit. Moreover, 
by comparing the SF to the flux of angular momentum across the black hole's 
event horizon, we shall prove that the conjecture (\ref{d-conjecture}) is 
satisfied also for strong fields. 

The SF on static charges in the spacetime of black holes was considered in
a number of works. The question to be asked then is the following: How is
the external force which is needed in order to keep the particle static
changed because of the self interaction of the particle? 
For the case of a static electric charge $q$ in the
spacetime of a Schwarzschild black hole of mass $M$, Smith and Will
\cite{smith-will-80} and Frolov and Zel'nikov
\cite{frolov-zelnikov-80,frolov-zelnikov-82} found that there is a
repelling inverse cubic force, which in the frame of a geodesic observer
who is momentarily at rest at the position of the charge $q$ is given by
\begin{equation}
F_{\hat\mu}=q^2\frac{M}{r^3}\,\delta^{\hat r}_{\hat\mu}\, , 
\label{smith-will}
\end{equation}
$r$ being the radial Schwarzschild coordinate. 
This force arises from the tail part of the full SF. The Schwarzschild
spacetime is Ricci flat, such that the coupling to Ricci curvature does
not contribute to the SF. Also, a static charge in Schwarzschild has zero
ALD force, such that the full SF (\ref{smith-will}) is given by just the
tail force. 

For the case of a static scalar charge (where the scalar field
is minimally-coupled) in Schwarzschild,
Wiseman \cite{wiseman-00} found the interesting result, that the SF is
zero (this was found earlier by Zel'nikov and Frolov
\cite{frolov-zelnikov-82} as a particular case in the more general study
of non-minimally coupled scalar fields). Since the SF again is given by
just the tail part (as Schwarzschild is Ricci flat, and the ALD force for
a static scalar charge is again zero), it turns out that the tail part of
the SF for a static scalar charge in Schwarzschild is zero. This vanishing
result is in some sense surprising: spacetime is curved in a non-trivial
way (which indeed leads to a non-zero tail force for an electric charge),
such that an exactly vanishing result for the tail force is not
intuitively expected. The following question then arises: Is this result
just a consequence of the particular symmetries imposed, which cause the
SF to take a zero value? When these symmetries are relaxed, does the zero
result for the SF persist, or do we find a non-zero result? Some aspects
of the symmetries were indeed relaxed in subsequent works. In Ref.\
\cite{burko-prl} the scalar charge was not considered anymore as static.
Instead, a scalar charge in uniform circular motion around Schwarzschild
was considered, and indeed a non-zero tail force, proportional to the
angular velocity squared, was found. This tail
force vanishes in the limit of zero angular velocity (a static particle). 
In Ref.\ \cite{shells-00} the horizon
condition was relaxed, and instead the spacetime was taken to be that of a
thin spherical shell. Again, a non-zero SF was found, which vanishes in
the limit that the radius of the shell approaches its Schwarzschild
radius. In these cases, too, the non-zero SF arises from the tail part of
the SF. We thus see that the zero result for the SF (and in particular for
the tail part of the SF) for the case of a static scalar charge in
Schwarzschild is just a degenerate case when more complicated cases are
considered, namely, a scalar charge in circular orbit, or a spacetime
which is Schwarzschild, but not that of a black hole. Is the result of a
zero SF for a static scalar charge outside a Schwarzschild black hole then
just an isolated result, or is it a particular case of a wider class of
spacetimes? 

We shall consider this question for a number of generalizations of the
Schwarzschild spacetime. First, we shall add to the black hole electric
charge, thus making it a Reissner-Nordstr\"{o}m black hole. The
Reissner-Nordstr\"{o}m spacetime is electrovac, and hence Ricci curved.
However, it turns out that the Ricci part of the SF for a static scalar
charge vanishes. Also the ALD part of the SF vanishes for that case. What
about the tail part of the SF? We find that the tail part of the SF is
zero, and so is the full SF. We thus see that by adding electric charge to
the black hole, we do not change the zero result for the SF. What about
adding angular momentum to the black hole? We study then the SF on a
static scalar charge in the spacetime of a Kerr black hole. Spacetime is
Ricci flat, but there is a local contribution to the SF from the ALD part
of the SF. This contribution is in the $\partial /\,\partial\varphi$
direction (in Boyer-Lindquist coordinates). (Incidentally, the force which
is required in order to hold the particle fixed in the absence of the SF
has components only in the $\partial /\,\partial r$ and $\partial
/\,\partial\theta$ directions.) When we consider the tail part of the SF,
we find that it is still zero. Namely, the full SF is given by just the
local, ALD part of the SF. We then consider the most general stationary,
axially symmetric black hole, by adding both angular momentum and electric
charge to the black hole, turning it into a Kerr-Newman black hole.
Spacetime now is Ricci curved, and there is a non-zero contribution to the
SF from the local Ricci-coupled part of the SF. There is also a
contribution from the local ALD part of the SF. However, when we consider
the full SF, we find that it equals just the sum of the two local terms,
such that the tail part of the SF is again zero. 

We thus conclude that the tail part of the SF on a static scalar charge
is zero for all stationary, axially symmetric black holes. The zero result
in the spacetime of a Schwarzschild black hole turns out then to be just
a particular case of a much wider class of spacetimes. In 
Schwarzschild also the local parts of the SF turn out to be zero, such
that the full SF vanishes. However, in more general spacetimes the
local terms are non-zero, such that there is a non-zero SF, but the
interesting (and the difficult to find) part of the SF is the tail part,
and it is zero for all stationary and axisymmetric black holes. 

We find that the full, regularized SF on a static scalar charge $q$ in the
spacetime of a Kerr-Newman black hole with mass $M$, spin parameter $a$,
and electric charge $Q$, is given in Boyer-Lindquist coordinates by
\begin{equation}
F_{\mu}^{\rm SF}=
\frac{1}{3}q^2a\Delta\sin^2\theta\frac{M^2-Q^2}
{(\Delta-a^2\sin^2\theta)^{5/2}\Sigma^{1/2}}\, \delta^{\varphi}_{\mu}\, .
\label{sf1}
\end{equation}     
Here, $q$ is the scalar charge of the particle, the horizon function
$\Delta=r^2-2Mr+a^2+Q^2$, and $\Sigma=r^2+a^2\cos^2\theta$. Equation
(\ref{sf1}) for the full, regularized SF acting on a static scalar
charge in the spacetime of a stationary, axisymmetric black hole is our
main result in this Paper.

One striking feature of the SF is that it diverges as the static limit
(the outer boundary of the ergosphere) is approached. This situation is
much different from that of a static electric charge in Schwarzschild,
given  by Eq.\ (\ref{smith-will}), where the SF is bounded. Also, the
ratio of the SF in that case to the regular force which is needed to be
applied in order to hold the charge static (in the absence of the SF)
tends to zero as the black hole's horizon is approached. Indeed, Smith and
Will found that the SF is just a tiny correction, which becomes important
only when the theory is no longer expected to be accurate, namely, when
quantum effects are needed to be considered \cite{smith-will-80}. (Smith
and Will found that, at its maximum at $r=3M$, the acceleration due to the
SF becomes comparable with the regular acceleration if the Schwarzschild
radius of the black hole is smaller than the classical electron radius,
and the distance of the electron from the horizon is smaller by two orders
of magnitude than the electron's Compton wave length.) 
In the strong field regime we thus find the SF acting on a scalar charge
held static in
the spacetime of a stationary, axisymmetric black hole to grow unboundedly  
as the static limit is approached. Note, that also the regular force
which is needed to
keep the particle static in the absence of the SF diverges in the same
limit. The ratio of the acceleration due to the SF, $a^{\rm SF}$, and the
regular acceleration,  $a^{\rm reg}$, for a Kerr black hole and on the
equatorial plane, is found to be 
\begin{equation}
\frac{a^{\rm SF}}{a^{\rm reg}}=\frac{1}{3}\,\frac{q^2}{\mu}\, a 
\frac{M}{r^2}\,\frac{1}{r-2M}\,
\end{equation}
where $q$ is the scalar charge of a particle with mass $\mu$, and $M$ and
$a$ are the mass and spin of the black hole, correspondingly. By the
accelerations of the left hand side we mean the corresponding magnitudes. 
As the
static limit is approached (on the equatorial plane the static limit is at
$r=2M$), this ratio diverges, which signifies that the acceleration due
to the SF becomes dominant over the regular acceleration. Also, there is a
finite value of $r$ for which the two accelerations become equal. When the
charge-to-mass ratio $q/\mu$ of the scalar particle is large, the two
accelerations become comparable at large distances from the black hole.

The organization of this paper is as follows. In Section \ref{form} we
derive the equations governing the field, and obtain the expression for
the SF. In Section \ref{weak} we derive the SF analytically in the weak  
field regime, and in Section \ref{sec:num} we evaluate the SF numerically
in the strong field regime. In Section \ref{sec:farfield} we derive the SF 
using the far field and balance arguments, and compare our results with those 
obtained in Sections \ref{weak} and \ref{sec:num} using the near field. 
Finally, in Section \ref{sf:prop} we discuss the properties of the SF.

\section{Formulation}\label{form}

Consider a static scalar charge in the spacetime of a stationary, 
axially symmetric, black hole, i.e.\ the Kerr-Newman black hole. 
By a static particle we mean here that the particle's spatial position is
fixed in Boyer-Lindquist coordinates. The background spacetime is
described by the Kerr-Newman metric, which in Boyer-Lindquist coordinates
assumes the form 
\begin{equation}
  ds^2=-\left( 1-\frac{2Mr-Q^2}{\Sigma}\right) dt^2 
-\frac{2a\sin^2 \theta (2Mr-Q^2)}
{\Sigma} dt d\varphi + {\Sigma \over \Delta}dr^2 + \Sigma d\theta^2 +
\frac{\zeta \sin^2 \theta}{\Sigma} d\varphi^2 \ ,
\label{kn-metric}
\end{equation}
where $\Sigma=r^2+a^2\cos^2 \theta$, $\Delta=r^2+a^2+Q^2-2Mr$, and 
$\zeta=(r^2+a^2)\Sigma+a^2(2Mr-Q^2)\sin^2\theta=(r^2+a^2)^2-\Delta
a^2 \sin^2\theta$. We use units in which $G=c=1$ throughout. Here $M$, $a$ and
$Q$ are respectively the mass, angular momentum per unit mass and
electric charge of the black hole. 

The linearized field equation for a minimally coupled, massless scalar field 
$\Phi$ is given by 
\begin{equation}
  \nabla_{\mu} \nabla^{\mu} \Phi(x^{\alpha}) = -4\pi \rho(x^{\alpha}) \ , 
\label{poisson}
\end{equation}
where $\nabla_{\mu}$ denotes covariant differentiation compatible with the 
metric (\ref{kn-metric}). The scalar charge density $\rho$ is given by 
\begin{equation}
  \rho(x^{\mu}) = q \int_{-\infty}^{\infty} d\tau \frac{\delta^4 [ x^{\mu}-
z^{\mu}(\tau)]}{\sqrt{-g}} \ .
\label{charge-density}
\end{equation}
Here $q$ is the particle's total scalar charge; $\tau$ is the proper time; 
$g=-\Sigma^2 \sin^2 \theta$ is the metric
determinant; and $z^{\mu}$ is the world line of the charge. We assume that the charge 
is placed at a position $(r_0,\theta_0,\varphi_0)$. To solve Eq.~(\ref{poisson}), 
we decompose $\rho$ and $\Phi$ into a sum over spherical harmonics 
$Y_{lm}(\theta,\varphi)$: 
\begin{eqnarray}
  \rho(r,\theta,\varphi) &=& q\sqrt{1-\frac{2Mr_0-Q^2}{\Sigma_0}}\, \frac{\delta(r-r_0)}
{\Sigma} \sum_{l=0}^{\infty} \sum_{m=-l}^l Y_{lm}^*(\theta_0,\varphi_0) 
Y_{lm}(\theta,\varphi) \label{rho-mode} \\
  \Phi(r,\theta,\varphi) &=& \sum_{l=0}^{\infty} \sum_{m=-l}^l \Phi^{lm}(r,\theta,\varphi) 
= \sum_{l=0}^{\infty} \sum_{m=-l}^l \phi^{lm}(r) Y_{lm}(\theta,\varphi) \ , 
\label{phi-mode}
\end{eqnarray}
where $\Sigma_0=r_0^2+a^2\cos^2\theta_0$. Substituting Eqs.~(\ref{rho-mode}) and 
(\ref{phi-mode}) into Eq.~(\ref{poisson}), we find
\begin{equation}
  \Delta \phi^{lm}_{,rr} + 2(r-M) \phi^{lm}_{,r} + \left[ \frac{m^2 a^2}{\Delta} 
-l(l+1) \right] \phi^{lm} = -4\pi q \sqrt{1-\frac{2Mr_0-Q^2}{\Sigma_0}}\,  
Y_{lm}^*(\theta_0,\varphi_0) \, \delta(r-r_0) \ ,
\label{mode-eq}
\end{equation}
where commas denote partial derivatives. The boundary conditions for $\Phi$ are 
that $\Phi$ vanishes as $r \rightarrow \infty$, and is regular on the future event 
horizon. 
Regularity of $\Phi$ on the future horizon is equivalent to the field
being derived from the retarded Green's function. (Regularity on the past
event horizon is similarly related to the advanced field.) 
The solution of Eq.~(\ref{mode-eq}) can be written as 
\begin{equation}
  \phi^{lm}(r) = \frac{\phi^{lm}_1(r_<)\, \phi^{lm}_2(r_>)}
{W_r[\phi^{lm}_1,\phi^{lm}_2](r_0)} S(r_0) Y_{lm}^*(\theta_0,\varphi_0) \ ,
\label{sol1}
\end{equation}
where 
\begin{equation}
  S(r_0)=-\frac{4\pi q}{\Delta_0} \sqrt{1-\frac{2Mr_0-Q^2}{\Sigma_0}} \ .
\label{Sr}
\end{equation}
Here $\Delta_0=r_0^2+a^2+Q^2-2Mr_0$, $r_>=\max(r,r_0)$, $r_<=\min(r,r_0)$,
$W_r[\phi^{lm}_1,\phi^{lm}_2](r_0)=\phi^{lm}_1(r_0) \phi^{lm}_{2,r}(r_0)
-\phi^{lm}_{2}(r_0) \phi^{lm}_{1,r}(r_0)$ is the Wronskian determinant  
evaluated at $r=r_0$; 
$\phi^{lm}_1$ and $\phi^{lm}_2$ are two independent solutions of the
homogeneous equation 
\begin{equation}
   \Delta \phi_{,rr} + 2(r-M) \phi_{,r} + \left[ \frac{m^2 a^2}{\Delta} 
-l(l+1) \right] \phi = 0 \ ,
\label{hom-eq}
\end{equation}
with $\phi^{lm}_2$ satisfying the boundary condition at infinity, and 
$\phi^{lm}_1$ chosen appropriately to make $\Phi^{lm}$ regular on the 
future event horizon.

Equation (\ref{hom-eq}) has three regular singular points at 
$r_+$, $r_-$, and at infinity, 
where $r_{\pm}=M\pm \sqrt{M^2-Q^2-a^2}$ are the outer and 
inner horizons of the black hole. We next move the regular singular points
of Eq.\ (\ref{hom-eq}) to $\pm 1,\infty$. This is done by the
transformation
\begin{equation}
  z(r)=\frac{2r-r_+-r_-}{r_+-r_-}\, .
\label{z}
\end{equation}
Equation (\ref{hom-eq}) then becomes 
\begin{equation}
  (1-z^2) \phi_{,zz} -2z\phi_{,z} + \left[ l(l+1)-\frac{\mu^2}{(1-z^2)}
\right] \phi =0 \ ,
\label{aleg-eq}
\end{equation}
which is the associated Legendre equation. 
Here the degree $\mu$ is purely imaginary and is given by 
\begin{equation}
  \mu=im\gamma \ \ \ , \ \ \ \gamma=\frac{a}{\sqrt{M^2-a^2-Q^2}} \ .
\label{mu}
\end{equation}
Two linearly independent solutions are $P_l^{\mu}(z)$ and $Q_l^{\mu}(z)$, 
the associated 
Legendre function of the first and second kinds~\cite{am-st},  
respectively. 
The functions $P_l^{\mu}(z)$ and $Q_l^{\mu}(z)$ can be expressed in terms of 
hypergeometric functions: 
\begin{eqnarray} 
  P_l^{\mu}(z) &=& \frac{1}{\Gamma(1-\mu)}\left(\frac{z+1}{z-1}\right)^{\mu/2} 
 {_2F_1} \left(-l,l+1;1-\mu;\frac{1-z}{2}\right) \label{Plmu} \\
  Q_l^{\mu}(z) &=& e^{i\mu \pi} 2^{-l-1}\sqrt{\pi}\, \frac{\Gamma(l+\mu+1)}
{\Gamma(l+{3\over 2})}\, z^{-l-\mu-1}(z^2-1)^{\mu/2}\, 
{_2F_1}\left(1+\frac{l+\mu}{2},
\frac{l+\mu+1}{2};l+{3\over 2};{1\over z^2}\right) \label{Qlmu} \ .
\end{eqnarray}
The hypergeometric function has the Gauss series representation 
\begin{equation}
  { _2F_1}(a,b;c;z)=\sum_{n=0}^{\infty} \frac{(a)_n (b)_n}{n!\,(c)_n} z^n \ \ \ 
|z|<1 \ ,
\label{hypfun}
\end{equation}
where Pochhammer's symbol is defined to be 
\begin{equation}
  (x)_0=1 \ \ \ \mbox{and} \ \ \  
(x)_n=x(x+1)\cdots (x+n-1) \ \ \ (n \geq 1) \ .
\end{equation}
Note that the hypergeometric function which appears in the 
$P_l^{\mu}(z)$ expression (\ref{Plmu}) is 
just a polynomial of order $l$, and the series 
expansion~(\ref{hypfun}) is valid 
even though the magnitude of the argument is greater than one.

As $r \rightarrow \infty$ ($z \rightarrow \infty$), $P_l^{\mu}(z) \sim z^l$ 
and $Q_l^{\mu}(z) \sim z^{-l-1}$. In order for $\Phi$ to vanish at infinity, 
we must have $\phi^{lm}_2(z)=Q_l^{\mu}(z)$ up to an arbitrary multiplicative 
factor [Note, from Eq.~(\ref{sol1}), that the value of $\phi^{lm}$ does not 
change if $\phi^{lm}_1$ or $\phi^{lm}_2$ is multiplied by a factor independent 
of $r$].

To determine $\phi^{lm}_1(z)$, we first consider the case $m=0$. The function 
$\Phi^{l0}(r,\theta,\varphi)=\phi(r) Y_{l0}(\theta,\varphi)$ is
independent of 
$\varphi$, such that the singularity of the coordinate $\varphi$ 
on the horizon does not 
complicate the analysis. The general solution of the homogeneous 
equation is the 
linear combinations of the Legendre functions $P_l(z)$ and $Q_l(z)$. On 
the horizon $z=1$, $Q_l$ diverges but $P_l$ remains finite, so we have 
$\phi^{l0}_1(z)=P_l(z)$ up to a multiplicative factor. When $m \neq 0$,
we find it more convenient to expand the general solution as a linear 
combination of $P_l^{\mu}(z)$ and $P_l^{-\mu}(z)$ instead of 
$P_l^{\mu}(z)$ and $Q_l^{\mu}(z)$. Note that $P_l^{-\mu}(z)$ is a solution 
of~(\ref{aleg-eq}) since the equation is invariant when $\mu$ is changed 
to $-\mu$. The two functions behave as 
\begin{equation}
  P_l^{\pm \mu}(z) \sim \frac{1}{\Gamma(1\mp \mu)}\left(\frac{z+1}{z-1}
\right)^{\pm \mu/2} 
= \frac{1}{\Gamma(1\mp \mu)}\left( \frac{r-r_-}{r-r_+}\right)^
{\pm i ma/(r_+-r_-)} \label{Plmu-asy}
\end{equation}
as $r \rightarrow r_+$. Both solutions oscillate near the horizon because of 
the coordinate singularity of $\varphi$ there. To remove this 
coordinate singularity,  we consider the coordinate
$\tilde{\varphi}(\varphi ,r)$ defined such that 
\[
  d \tilde{\varphi}=d\varphi+\frac{a}{\Delta}dr \ .
\]
Upon integration, we have
\begin{equation}
  \tilde{\varphi} = \varphi + \int \frac{a}{\Delta} dr 
= \varphi + \frac{a}{r_+-r_-}
\ln \left( \frac{r-r_+}{r-r_-} \right) \ 
\label{phi-inkn}
\end{equation}
plus an integration constant which we set equal to zero without loss of
generality (it only relates to the origin of the coordinate
$\tilde{\varphi}$.) 
Hence 
\[
  e^{im\tilde{\varphi}} = \left( \frac{r-r_-}{r-r_+}\right)^{ima/(r_+-r_-)}
e^{im\varphi} \ .
\]
In terms of the $(r,\theta,\tilde{\varphi})$ coordinates (the ingoing
Kerr-Newman coordinates), the two functions
$P_l^{\mu}(z)Y_{lm}(\theta,\varphi)$ and 
$P_l^{-\mu}(z)Y_{lm}(\theta,\varphi)$ become
\[ 
  \frac{1}{\Gamma(1-\mu)} Y_{lm}(\theta,\tilde{\varphi}) \left( \frac{r-r_-}
{r-r_+}
\right)^{2ima/(r_+-r_-)} 
\ \ \ \mbox{and} \ \ \ \frac{1}{\Gamma(1+\mu)} 
Y_{lm}(\theta,\tilde{\varphi})
\]
as $r\rightarrow r_+$ ($z\rightarrow 1$), respectively. (When regularity
of $\Phi$ on the past event horizon is required, the appropriate
coordinate transformation is given by  
$d \tilde{\varphi}=d\varphi-\frac{a}{\Delta}dr$.) The first
expression still
oscillates near the horizon while the second one is regular. Combining with the 
$m=0$ case, we conclude that $\phi^{lm}_1(z)=P_l^{-\mu}(z)$  up to 
a multiplicative factor.

Note that each mode of the potential $\Phi^{lm}(r,\theta,\varphi)=\phi^{lm}(r)
Y_{lm}(\theta,\varphi)$ is complex in general. However, it is easy to show that 
$\Phi^{l,-m}=(\Phi^{lm})^*$, such that the $m$-sum in Eq.~(\ref{phi-mode}) 
is real and so is the scalar field $\Phi$.

The modes of the bare SF, $f^l_{\mu}$, are given by
\begin{equation}
  f^l_{\mu} = \sum_{m=-l}^l q \nabla_{\mu} \Phi^{lm}(r_0,\theta_0,\varphi_0) \ .
\label{fl}
\end{equation}
It follows from Eqs.~(\ref{phi-mode}) 
and (\ref{sol1}) that $\Phi^{lm}_{,\theta}$ and $\Phi^{lm}_{,\varphi}$ are
continuous at the position of the charge, but $\Phi^{lm}_{,r}$ is not. Hence
$f^l_r$ is not uniquely defined. However, the MSRP guarantees that the
regularized SF does not depend on which derivative [$\Phi^{lm}_{,r}(r_0^+)$,
$\Phi^{lm}_{,r}(r_0^-)$ (where $r_0^{\pm}$ are the one sided limits of
$r\to r_0$ from above or below, correspondingly) or the average of the
two] we use~\cite{barack-00}. In practice, we use the average derivative, i.e., 
we define
\begin{equation}
  f^l_r=\frac{1}{2}\sum_{m=-l}^l q [\Phi^{lm}_{,r}(r_0^+)+
\Phi^{lm}_{,r}(r_0^-)] \ .
\end{equation}                     
Explicitly, we find that 
\begin{equation}
f^l_r(r_0,\theta_0,\varphi_0)=
\frac{1}{2}\sum_{m=-l}^{l}\frac{qS(r_0)}{W_{r}[\phi^{lm}_1,\phi^{lm}_2](r_0)}
\left[\phi^{lm}_1(r_0)\phi^{lm}_{2,r}(r_0)+
\phi^{lm}_{1,r}(r_0)\phi^{lm}_{2}(r_0)\right]Y_{lm}^*(\theta_0,\varphi_0)
Y_{lm}(\theta_0,\varphi_0)\, , 
\label{fr-formal}
\end{equation}
\begin{equation}
f^l_{\theta}(r_0,\theta_0,\varphi_0)=
\sum_{m=-l}^{l}\frac{qS(r_0)}{W_{r}[\phi^{lm}_1,\phi^{lm}_2](r_0)}
\phi^{lm}_1(r_0)\phi^{lm}_{2}(r_0)Y_{lm}^*(\theta_0,\varphi_0)
\,\partial_{\theta}Y_{lm}(\theta_0,\varphi_0)\, ,
\label{fth-formal}
\end{equation}          
and
\begin{equation}
f^l_{\varphi}(r_0,\theta_0,\varphi_0)=
\sum_{m=-l}^{l}\frac{qS(r_0)}{W_{r}[\phi^{lm}_1,\phi^{lm}_2](r_0)}
\phi^{lm}_1(r_0)\phi^{lm}_{2}(r_0)Y_{lm}^*(\theta_0,\varphi_0)
\,\partial_{\varphi}Y_{lm}(\theta_0,\varphi_0)\, .
\label{fphi-formal}
\end{equation}     
Typically, $\sum_l f^l_{\mu}$ diverges when summed na\"{\i}vely. 
We then use MSRP to regularize the SF as outlined in Section I. 
Recall that the MSRP parameters $a_{\mu},b_{\mu}$, and $c_{\mu}$ can be 
determined by two independent methods. Specifically, they can be found by either 
(i) study of the large-$l$ behavior of the individual modes of the bare SF,
or by (ii) a local analysis of the Green function. In this paper, we shall study
the regularization parameters using the former method. (It is hard to
apply the latter method here because of the following reason. 
Any time-dependent evolution of the wave equation in the spacetime of
a rotating black hole has to handle mode couplings, and the Green's function 
is obviously time dependent \cite{barack-99}.)

We shall carry out the regularization procedure analytically 
in Section~\ref{weak} to study the SF in the weak field regime, 
and then numerically in Section~\ref{sec:num} in the strong field regime.

\section{Computation of Self-Force in the Weak Field Regime}\label{weak}

In this Section, we consider the case for which 
the scalar charge is far away from 
the black hole, i.e.\ $r_0 \gg M$ [$z_0=(2r_0-r_+-r_-)/(r_+-r_-) \gg 1$]. We 
can thus expand $f_{\mu}^l$ in Eq.~(\ref{fl}) in powers of $r_0^{-1}$. 

The large $z_0$ expansion of $\phi^{lm}_1$ and $\phi^{lm}_2$ can be carried out 
by using Eqs.~(\ref{Plmu}) and (\ref{Qlmu}). Specifically, we choose 
\begin{eqnarray}
  \phi^{lm}_1(z) &=& \frac{2^l l!}{(2l)!}\Gamma(1+\mu+l) P_l^{-\mu}(z) = 
 \frac{2^l l!}{(2l)!}\frac{\Gamma(1+\mu+l)}{\Gamma(1+\mu)} 
\left( {z+1 \over z-1}\right)^{-\mu/2} { _2F_1}\left(-l,l+1;1+\mu;{1-z\over 2} 
\right) \\
  \phi^{lm}_2(z) &=& \frac{e^{-i\mu \pi} 2^{l+1}}{\sqrt{\pi}}
\frac{\Gamma \left( l+{3\over 2}\right)}{\Gamma(l+\mu+1)}Q_l^{\mu}(z) = 
z^{-l-\mu-1} (z^2-1)^{\mu/2} { _2F_1}\left(1+\frac{l+\mu}{2},\frac{l+\mu+1}{2}; 
l+{3\over 2}; {1\over z^2} \right) \ .
\end{eqnarray}
It can be shown, from Eq.~(\ref{hypfun}), that the above expressions are 
equivalent 
to 
\begin{eqnarray}
  \phi^{lm}_1(z) &=& z^l \left(1+{1\over z}\right)^{-\mu/2} \left(1-{1\over z}
\right)^{l+\mu/2} \sum_{n=0}^l \frac{(-l)_n (-\mu-l)_n}{n!\,(-2l)_n} \left( 
\frac{2}{1-z}\right)^n \label{phi1} \\
  \phi^{lm}_2(z) &=& {1\over z^{l+1}} \left( 1-{1\over z^2}\right)^{\mu/2} 
\sum_{n=0}^{\infty} \frac{\left(1+{l+\mu \over 2}\right)_n \left({l+\mu+1 \over 
2}\right)_n}{n! \left( l + {3\over 2}\right)_n} {1\over z^{2n}} \ .
\label{phi2}
\end{eqnarray}
Up to this point, no approximation has been made, but Eqs.~(\ref{sol1}), 
(\ref{phi1}), (\ref{phi2}) and (\ref{fl}) give us a convenient way to expand
$f^l_{\mu}$ in powers of $r_0^{-1}$. 
Hereafter, we evaluate all quantities at the position of the particle. To 
simplify the notation we shall assume that the scalar charge is placed at 
$(r,\theta,\varphi)$ and drop all the subscripts ``0''. 

We first consider the expansions for $f_r^l$ and $f_{\theta}^l$. We shall compute 
$f_r^l$ up to the order $r^{-7}$ and $f^l_{\theta}$ up to the order $r^{-6}$.
Straightforward calculations give 
\begin{eqnarray}
  f_r^l &=& \frac{q S(r)}{2}\left \{ {X_l(0,\theta)\over 2l+1}+\frac{2[l(l+1)X_l(0,\theta)
+3\gamma^2 X_l(2,\theta)]}
{(2l-1)(2l+1)(2l+3)}\frac{1}{z^2}+
\frac{6}{(2l-3)(2l-1)(2l+1)(2l+3)(2l+5)}\times
\right. \cr
 & & \left. [l(l+1)(l^2+l-3)X_l(0,\theta)+2(3l^2+3l-5)\gamma^2 X_l(2,\theta)
+5\gamma^4 
X_l(4,\theta)]\frac{1}{z^4} + 
O\left({1\over z^6}\right) \right \} \label{fr1} \\
  f_{\theta}^l &=& -q\sqrt{M^2-a^2-Q^2}S(r)\left \{ \frac{2\gamma^2\xi_l(2,\theta)}
{(2l-1)(2l+1)(2l+3)}\frac{1}{z}+\frac{2[3\gamma^4\xi_l(4,\theta)+2l(l+1)\gamma^2
\xi_l(2,\theta)]}
{(2l-3)(2l-1)(2l+1)(2l+3)(2l+5)}\frac{1}{z^3} 
+ O\left( {1\over z^5}\right)\right \} \ ,
\label{fth1}
\end{eqnarray}
where 
\begin{eqnarray}
  X_l(p,\theta) &=& \sum_{m=-l}^l m^p Y_{lm}^*(\theta,\varphi) 
Y_{lm}(\theta,\varphi) \\ 
  \xi_l(p,\theta) &=& \sum_{m=-l}^l m^p Y_{lm}^*(\theta,\varphi) 
\partial_{\theta} 
Y_{lm}(\theta,\varphi) \ .
\end{eqnarray}
In writing Eqs.~(\ref{fr1}) and (\ref{fth1}), we have used the results that 
$\xi_l(0,\theta)=0$ and $X_l(p,\theta)=\xi_l(p,\theta)=0$ 
if $p$ is a positive odd 
integer. We leave the detailed calculations of the two functions 
$X_l(p,\theta)$ and $\xi_l(p,\theta)$ to  
Appendix~\ref{app-a}. Substituting the expressions of $X_l$ and $\xi_l$
from Appendix~\ref{app-a} [Eqs.\ (\ref{a17})--(\ref{a22})], we obtain 
\begin{eqnarray}
  f_r^l &=& q\frac{S(r)}{8\pi} \left \{ 1 + \frac{2l(l+1)}{(2l-1)(2l+3)}
\left[ 1+\frac{3\gamma^2}{2}\sin^2\theta\right] \frac{1}{z^2}
+ \frac{6l(l+1)}
{(2l-3)(2l-1)(2l+3)(2l+5)}\times \right. \cr
 & & \left. \left[ (l^2+l-3)+(3l^2+3l-5)\gamma^2 \sin^2 \theta+
\frac{5}{2}\gamma^4 \sin^2 
\theta+\frac{15}{8} \gamma^4 (l-1)(l+2)\sin^4 \theta \right] \frac{1}{z^4} 
+ O \left({1\over z^6}\right) \right \} \label{flr1} \\
  f_{\theta}^l &=& -q\frac{S(r)}{4\pi}\sqrt{M^2-a^2-Q^2}\gamma^2 \sin
\theta \cos \theta \left \{ 
\frac{l(l+1)}{(2l-1)(2l+3)}\frac{1}{z}
+\frac{2l(l+1)}{(2l-3)(2l-1)(2l+3)(2l+5)}\times
\right. \cr 
& & \left. \left[ l(l+1)+{3\over 2}\gamma^2 +{9\over 4}(l-1)(l+2)
\gamma^2 \sin^2 
\theta \right] \frac{1}{z^3}+ O\left({1\over z^5}\right) \right \} \ \ .
\label{flth1}
\end{eqnarray}
For large values of $l$ we find 
\begin{eqnarray}
  f^{l\gg 1}_r &=& \frac{q S(r)}{8\pi} \left[ 
1 + {1\over 2} \left(1+ \frac{3\gamma^2}{2}\sin^2 \theta \right)
\frac{1}{z^2} + 
{3\over 8} \left( 1 + 3\gamma^2 \sin^2 \theta + 
{15\over 8}\gamma^4 \sin^4 \theta \right)\frac{1}{z^4}
+ O \left( {1\over z^6}\right)\right] + O \left({1\over l^2}\right)
\label{flrlim} \\ 
  f^{l\gg 1}_{\theta} &=& -\frac{q S(r)}{16\pi} 
\sqrt{M^2-a^2-Q^2}\, \gamma^2 \sin \theta \cos \theta 
\left[{1\over z}+
{1\over 2} \left(1+{9\over 4}\gamma^2 \sin^2 \theta \right)
\frac{1}{z^3} + O\left({1\over z^5}\right) \right] 
+ O \left( {1\over l^2}\right)   \ .
\label{flthlim}
\end{eqnarray}
Hence we find that the MSRP parameters $a_r=c_r=0$ up to the order of 
$r^{-7}$ (i.e., any deviation of $a_r$ or $c_r$ from zero is of order 
$r^{-8}$ or higher) [Note that $S(r)$ gives an extra factor $r^{-2}$],  
$a_{\theta}=c_{\theta}=0$ up to the order of $r^{-6}$ 
(i.e., any deviation of $a_{\theta}$ or $c_{\theta}$ from zero is of order
$r^{-7}$ or higher), 
and $b_r$ and $b_{\theta}$ are given from Eqs.~(\ref{flrlim}) 
and (\ref{flthlim}):
\begin{equation}
b_{r}=\frac{q S(r)}{8\pi} \left[
1 + {1\over 2} \left(1+ \frac{3\gamma^2}{2}\sin^2 \theta \right)\frac{1}{z^2} 
+{3\over 8} \left( 1 + 3\gamma^2 \sin^2 \theta +
{15\over 8}\gamma^4 \sin^4 \theta \right) \frac{1}{z^4} 
+ O \left( {1\over z^6}\right) \right] \label{br}
\end{equation}
and 
\begin{equation} 
b_{\theta}=
-\frac{q S(r)}{16\pi}
\sqrt{M^2-a^2-Q^2}\, \gamma^2 \sin \theta \cos \theta 
\left[{1\over z}+
{1\over 2} \left(1+{9\over 4}\gamma^2 \sin^2 \theta \right)\frac{1}{z^3}
+ O\left({1\over z^5}\right) \right]  \, .
\label{btheta}
\end{equation}
[For the case for which the particle is on the polar axis of the Kerr-Newman
black hole we find that 
\begin{equation}
b_{r}^{\rm axis}=-\frac{q^2}{2r^2}\, \frac{1-M/r}{1-2M/r+(a^2+Q^2)/r^2}
\left(1+\frac{a^2}{r^2}\right)^{-1/2}\, \delta_{\mu}^{r}\, . 
\end{equation}
We have checked the latter expression numerically in the strong
field regime, and found complete agreement.] 
Assuming the conjecture~(\ref{full-SF}), the $r$-component 
of the regularized SF is then calculated by 
subtracting Eq.~(\ref{br}) from Eq.~(\ref{flr1}) and then
summing over $l$. The $\theta$-component is evaluated similarly from Eqs.\  
(\ref{btheta}) and (\ref{flth1}). The results are
\begin{eqnarray}
  F_r &=& \sum_{l=0}^{\infty} (f^l_r-b_r) = 0+ O\left({1\over r^8}\right) \\
  F_{\theta} &=& \sum_{l=0}^{\infty} (f^l_{\theta}-b_{\theta}) = 0+ O
\left({1\over r^7}\right) \ ,
\end{eqnarray}
where we have used the fact that for any integer $k$ 
\begin{equation}
  \sum_{l=0}^{\infty} \left[ \frac{1}{2(l-k)+1}-\frac{1}{2(l+k)+1}\right]=0\, . 
\end{equation}
Hence we conclude that any non-zero orthonormal 
$r$ and $\theta$ components of the SF
($F_{\hat{r}}$ and $F_{\hat{\theta}}$) are of order $r^{-8}$ or higher.
In the next section, we present strong numerical evidence to suggest that 
$F_r$ and $F_{\theta}$ are actually zero wherever the location of the charge.

The $\varphi$-component of the SF can also be computed in the same 
way. However, we find a better method to do the calculation, which is 
described in detail in Appendix~\ref{app-b}. We find that 
\begin{equation}
  f^l_{\varphi}=-\frac{q S(r)}{z^{2l}} a  
\left(1-{1\over z^2}\right) \sum_{m=-l}^l 
m^2 \left[ \prod_{j=1}^l (m^2\gamma^2+j^2)\right] 
\left[ \frac{K_l^{\mu}(z)}{(2l+1)!!}\right]^2 Y^*_{lm}(\theta,\varphi)
Y_{lm}(\theta,\varphi) \ ,
\label{fphi}
\end{equation}
where $(2l+1)!!=1\cdot 3\cdot 5 \cdots (2l+1)$ and 
\begin{equation}
  K^{\mu}_l(z)=\left(1-{1\over z^2}\right)^{\mu/2} { _2F_1}\left( 
1+\frac{l+\mu}{2},\frac{l+\mu+1}{2};l+{3\over 2};{1\over z^2}\right) \ .
\end{equation}
It follows from Eq.~(\ref{fphi}) that the leading term of $f^l_{\varphi}$ 
is of order $r^{-2l-2}$. So when we make an asymptotic expansion by keeping terms up to
$r^{-N}$, only finite number of terms with $l\leq (N-2)/2$ contribute. 
In other words, up to the order $r^{-N}$, $f^l_{\varphi}=0$ when $l>(N-2)/2$, 
such that the MSRP parameters $a_{\varphi}=b_{\varphi}=c_{\varphi}=0$. 
In the next Section, we show numerically that $f^l_{\varphi}$ decreases 
exponentially with increasing $l$, which also suggests that no regularization 
is needed to compute $F_{\varphi}$. Expanding Eq.~(\ref{fphi}) in powers 
of $r^{-1}$, we find that  
\begin{equation}
  F_{\varphi}=\sum_{l=0}^{\infty} f^l_{\varphi} =\frac{q^2}{3}a\sin^2\theta
\frac{M^2-Q^2}{r^4}
\left \{ 1+3{M\over r}+{1\over 2}[3(5M^2-Q^2)+2a^2(1-3\cos^2\theta)]
{1\over r^2} + O\left({1\over r^3}\right)
\right \}  \ .
\label{exp-1}
\end{equation}
The leading order term of this expansion agrees with the result given in 
Ref.\ \cite{galtsov-82} for the case where $Q=0$. This agreement implies that 
we use here the correct parameter $d_{\varphi}$, and that the conjecture 
(\ref{d-conjecture}) holds for the case studied here. 
Note that this conclusion is valid only for the weak-field regime of a 
Kerr spacetime. In the next sections we bring evidence that this 
conjecture holds also for cases for which $Q\ne 0$, and also in strong fields.

\section{Computation of the self force in the strong field 
regime}\label{sec:num}

Next, we compute the SF in the strong-field regime. To do so, we use the
general expressions (\ref{fr-formal})--(\ref{fphi-formal}) 
for the modes of the field, and compute them numerically. In order to find the
regularized SF, we shall use MSRP. As discussed above, we determine the MSRP parameters
$a_{\mu},b_{\mu}$, and $c_{\mu}$ by studying the large-$l$ behavior of the
individual modes of the bare SF. We shall, however, check our numerical
results for the regularization parameters in the spherically symmetric
limit, where the regularization parameters are known analytically
\cite{barack-00}. This is done in subsection \ref{subsec:rn},
where we study the case of a Reissner-Nordstr\"{o}m black hole. Then, in 
subsection  \ref{subsec:kerr} we consider the case of a Kerr black hole,
and in subsection \ref{subsec:kn} we consider the case of a Kerr-Newman
black hole. We shall also compare our results for the strong field with 
the weak-field approximation of Section \ref{weak}. 

\subsection{Reissner-Nordstr\"{o}m}\label{subsec:rn}
Our main goal in this Paper is to
study the SF on a static scalar charge in the spacetime of a rotating
black hole. We shall apply our computation also for the
spherically symmetric, electrically charge, Reissner-Nordstr\"{o}m black
hole for two reasons. First, as we already noted, the SF acting on a static 
scalar charge in the spacetime of a
Reissner-Nordstr\"{o}m black hole has not been calculated yet. Our results
in this subsection are thus new. 
Second, it will allow us to check our numerical code for a case
where the MSRP parameters are known analytically. (Our code does not
assume spherical symmetry, thus its computation of the angular dependence
of functions is non-trivial.)  

\begin{figure}
\epsfig{file=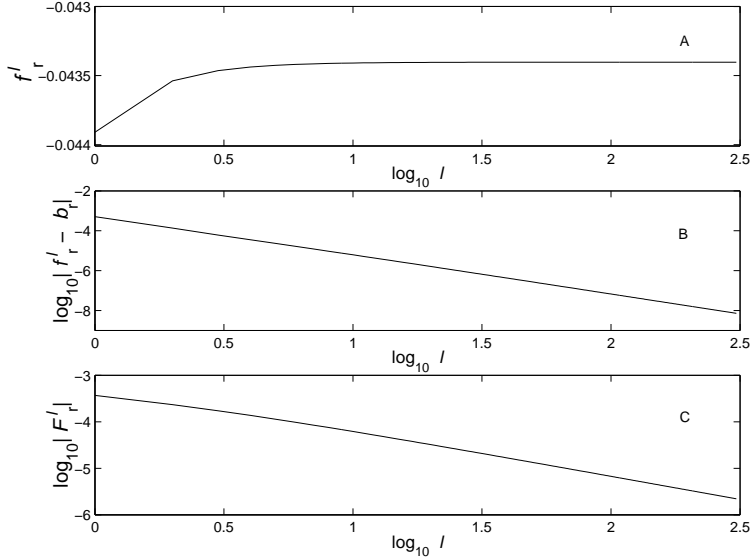,width=10cm,angle=0}
\caption{Self force on a static scalar charge in Reissner-Nordstr\"{o}m.
Upper panel (A): $f^{l}_r$ as a function of $l$. 
Middle panel (B): The difference $f^{l}_r-b_r^{\rm RN}$ as a function of 
$l$. 
Lower panel (C): $F^l_r$  as a function of $l$.
The data presented here correspond to the parameters:
$r=4M$, $a=0$, $Q=0.8M$, and $M=1$. For the actual computation we took
$\theta=\pi/4$.}
\label{rn}
\end{figure}

It is clear from symmetry  considerations that all the
azimuthal components vanish, such that we check below only the radial
component of the SF. In Reissner-Nordstr\"{o}m, it was found
analytically in Ref.\ \cite{barack-00} that for a static scalar charge 
\begin{equation}
b_r^{\rm RN}=-\frac{q^2}{2r^2}\,\frac{1-M/r}{1-2M/r+Q^2/r^2}\, ,
\label{br-rn}
\end{equation}
and $a_r^{\rm RN}=0=c_r^{\rm RN}$. Also, $d_r^{\rm RN}=0$. 
Figure \ref{rn} displays our results for a Reissner-Nordstr\"{o}m black
hole of mass $M=1$ and electric charge $Q=0.8M$, 
for a particle at $r_0=4M$. 
In Fig.\ \ref{rn}(A) we present the modes of the covariant
radial component of the bare SF, $f^l_r$, for a Reissner-Nordstr\"{o}m
spacetime. These modes appear to approach a constant value in the
large-$l$ limit. In order to check whether this limiting value coincides
with $b_r^{\rm RN}$, we plot in Fig.\ \ref{rn}(B) the difference between
$f^l_r$ and $b_r^{\rm RN}$ as a function of $l$. As for
Reissner-Nordstr\"{o}m it was shown that the MSRP parameter $d_r^{\rm
RN}=0$, and also the ALD and Ricci-curvature parts of the SF vanish, the 
conjecture (\ref{d-conjecture}) holds. Hence  
this difference is just the modes of the regularized
full SF. 
We find this difference to scale like $l^{-2}$ for large 
values of $l$. This result is in full
accord with the analytical results for the MSRP parameters. We also
compute the regularized full SF. In Fig.\ \ref{rn}(C) we plot 
$F^l_r=\sum_{j=0}^{l}(f^j_r-b_r^{\rm RN})$ as a function of $l$. We find
that $F^l_r$ behaves like $l^{-1}$ in the large-$l$ limit, such that
$F^l_r\to 0$ as $l\to\infty$. We infer that $F_r\equiv F_r^{l\to\infty}=0$
for a static scalar charge in Reissner-Nordstr\"{o}m.  
Similar results were obtained for other choices of the parameters. 
We infer that both the tail, non-local part of the SF, 
and the local two terms of the SF vanish separately for a static scalar 
charge outside a Reissner-Nordstr\"{o}m black hole.  
This generalizes the known result of a
zero SF on a static scalar charge in Schwarzschild to
Reissner-Nordstr\"{o}m.

\subsection{Kerr}\label{subsec:kerr}
When the black hole is endowed with non-zero spin, we no longer have
analytical results for the MSRP parameters $a_{\mu},b_{\mu}$ and
$c_{\mu}$. As noted above, these parameters can be found, however, from
the large-$l$ behavior of the modes of the bare SF. Nevertheless, the MSRP
parameter $d_{\mu}$ cannot be found by studying the modes of the bare SF
alone, and a local analysis of the Green's function is necessary in order
to determine it. Recently, a conjecture about the MSRP parameter $d_{\mu}$
was formulated \cite{barack-00,burko-amaldi}. According to this
conjecture, the MSRP parameter $d_{\mu}$ equals the sum of the two local
terms in the full regularized SF, as given by Eq.\ (\ref{d-conjecture}). 
In the Kerr spacetime, which is Ricci
flat, $d_{\mu}$ equals then just the ALD force. In the following we shall
use this conjecture. The ALD part of the SF is given by
\begin{equation}
F_{\mu}^{\rm ALD}=
\frac{1}{3}q^2aM^2\Delta\frac{\sin^2\theta}{(\Delta-a^2\sin^2\theta)^
{5/2}\Sigma^{1/2}}\, \delta^{\varphi}_{\mu}\, .
\label{kerr:ald}
\end{equation}
We thus conjecture that $d_{\mu}=F_{\mu}^{\rm local}$. Also, because
spacetime is Ricci flat, $F_{\mu}^{\rm local}=F_{\mu}^{\rm ALD}$.

\begin{figure}
\epsfig{file=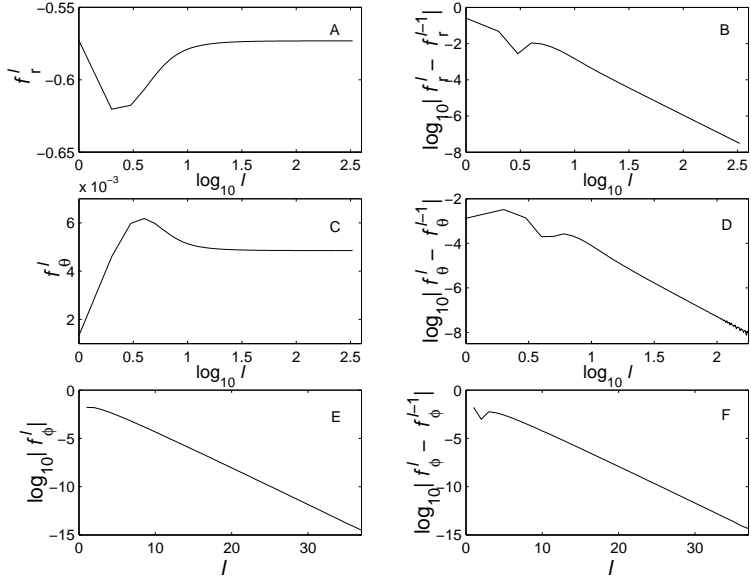,width=10cm,angle=0}
\caption{Behavior of the individual modes of the self force for a
static scalar charge in the spacetime of a Kerr black hole. Panels (A),
(C), and (E): The $r$, $\theta$, and $\varphi$ covariant components of the
individual modes of the self force, respectively
(i.e., $f^{l}_{r}$, $f^{l}_{\theta}$, and $f^{l}_{\varphi}$), as functions
of the mode number $l$. Panels (B), (D), and (F): The $r$, $\theta$,
and $\varphi$ covariant components of the difference between
two consecutive modes of the self force, respectively
(i.e., $f^{l}_{r}-f^{l-1}_{r}$, $f^{l}_{\theta}-f^{l-1}_{\theta}$, and
$f^{l}_{\varphi}-f^{l-1}_{\varphi}$), as functions
of the mode number $l$. The parameters for the data presented here are:
$r=2.2M$, $\theta=\pi/4$, $a=0.2M$, $Q=0$, and $M=1$.}
\label{modeskerr}
\end{figure}

Figure \ref{modeskerr} displays the behavior of the
individual modes of the bare SF for a static scalar charge in the
spacetime of a Kerr black hole (i.e., the scalar charge has fixed
Boyer-Lindquist coordinates $r,\theta$ and $\varphi$.) 
We choose here the 
parameters $r=2.2M$, $\theta=\pi/4$, $a=0.2M$, $Q=0$, and $M=1$, but
similar results were found also for other choices. We find that
$f^{l}_{r}$ and $f^{l}_{\theta}$ approach constants as $l\to\infty$, and
that the difference between two consecutive modes, $f^{l}_{r}-f^{l-1}_{r}$
and $f^{l}_{\theta}-f^{l-1}_{\theta}$, behave like $l^{-3}$ in the
large-$l$ limit. This behavior implies that the MSRP parameters $a^{\rm
Kerr}_{r},a^{\rm Kerr}_{\theta},c^{\rm Kerr}_{r}$, and 
$c^{\rm Kerr}_{\theta}$ vanish. (Non-zero $a^{\rm Kerr}_{\mu}$ implies
linear growth of the modes with the mode number $l$, and non-zero 
$c^{\rm Kerr}_{\mu}$ implies that the difference between two consecutive
modes behaves like $l^{-2}$. Moreover, 
non-zero $c^{\rm Kerr}_{\mu}$ threatens the applicability of MSRP, 
as it implies a divergent $d^{\rm Kerr}_{\mu}$ \cite{barack-00}.) 
Recall that we have defined $a_{\mu}$ in the averaged
sense. The corresponding ``one sided'' values are not zero, in
general. We also find that $f^{l}_{\varphi}$ decays exponentially with
$l$ for large values of $l$, which suggests that 
$a^{\rm Kerr}_{\varphi}=0=c^{\rm Kerr}_{\varphi}$, and, in addition, also 
$b^{\rm Kerr}_{\varphi}=0$. The vanishing of $b^{\rm Kerr}_{\varphi}$ is
in accord with our results in the weak-field expansions in Section \ref{weak}. 

In order to compute the regularized SF we have to confront the difficulty
of not having exact expressions for the MSRP parameter $b_{\mu}^{\rm
Kerr}$. However, $b_{\mu}^{\rm Kerr}$ is nothing but the limit as
$l\to\infty$ of $f^{l}_{\mu}$. We can thus approximate $b_{\mu}^{\rm
Kerr}$ by studying the large-$l$ behavior of $f^{l}_{\mu}$, and
extrapolate to infinite value of the mode number (e.g., through
Richardson extrapolation). In practice, we can
approximate $b_{\mu}^{\rm Kerr}$ by simply taking the mode of the bare SF
with a mode number $L$ much larger than the mode $l$ up to which we sum
over the
modes to obtain the full, summed-over-modes, SF. That is, we compute the
regularized SF according to
\begin{eqnarray}
F_{\mu}&=&\sum_{j=0}^{\infty}(f_{\mu}^{j}-f_{\mu}^{\infty})\equiv 
\sum_{j=0}^{l}(f_{\mu}^{j}-f_{\mu}^{\infty})+{\cal R}^{l+1}_{\mu}
\label{R-def} \\
&\equiv &
\sum_{j=0}^{l}(f_{\mu}^{j}-f_{\mu}^{L})+{\cal R}^{l+1}_{\mu}+
{\cal E}^{L}_{\mu} \, .
\label{F}
\end{eqnarray}
The functions ${\cal R}^{l+1}_{\mu}$ and ${\cal E}^{L}_{\mu}$ are defined
by Eqs. (\ref{R-def}) and (\ref{F}), respectively. Now, by definition
\begin{eqnarray}
{\cal R}^{l+1}_{\mu}&=&\sum_{j=l+1}^{\infty}(f_{\mu}^{j}-f_{\mu}^{\infty})
\nonumber \\
&=& \sum_{j=l+1}^{\infty} \left[ \frac{x_{\mu}}{j^2}+O(j^{-3})\right]\, ,
\label{r-ap}
\end{eqnarray}
where $x_{\mu}$ is the coefficient of the $l^{-2}$ term in the $l^{-1}$
expansion of $f_{\mu}^{l}$. In evaluating ${\cal R}^{l+1}_{\mu}$ we shall
drop the $O(j^{-3})$ term in Eq.\ (\ref{r-ap}). This introduces an error which 
reduces like $l^{-2}$ as $l$ grows.  By definition, the
function ${\cal E}^{L}_{\mu}= 
\sum_{j=0}^{l}(f_{\mu}^{L}-f_{\mu}^{\infty})$, and as inside the sum we
just have a $j$-independent expression, we find that 
${\cal E}^{L}_{\mu}=(l+1)(f_{\mu}^{L}-f_{\mu}^{\infty})$.

Taking $L\gg l\gg 1$, we approximate $F_{\mu}$ in practice by 
\begin{equation}
F_{\mu}^l\approx \sum_{j=0}^{l}(f_{\mu}^{j}-f_{\mu}^{L}) .
\end{equation}
The error associated with this approximation has two contributions, i.e., 
${\cal R}^{l+1}_{\mu}$ and ${\cal E}^{L}_{\mu}$. We evaluate  
${\cal R}^{l+1}_{\mu}\approx
x_{\mu}\psi^{(1)}(l+1)\approx x_{\mu}/l$, where
$\psi^{(1)}(z)$ is the trigamma function. We also evaluated 
${\cal E}^{L}_{\mu}\approx lx_{\mu}/L^2$. Note, that $x_{\mu}$ can be
evaluated from the difference between two consecutive modes, i.e., 
$x_{\mu}\approx l^3(f^{l-1}_{\mu}-f^{l}_{\mu})/2$ (when $l$ is large
enough). We find that  
${\cal E}^{L}_{\mu}/{\cal R}^{l+1}_{\mu}\approx (l/L)^2$, such that
when $L\gg l$ the overall error is dominated by ${\cal R}^{l+1}_{\mu}$. 

\begin{figure}
\epsfig{file=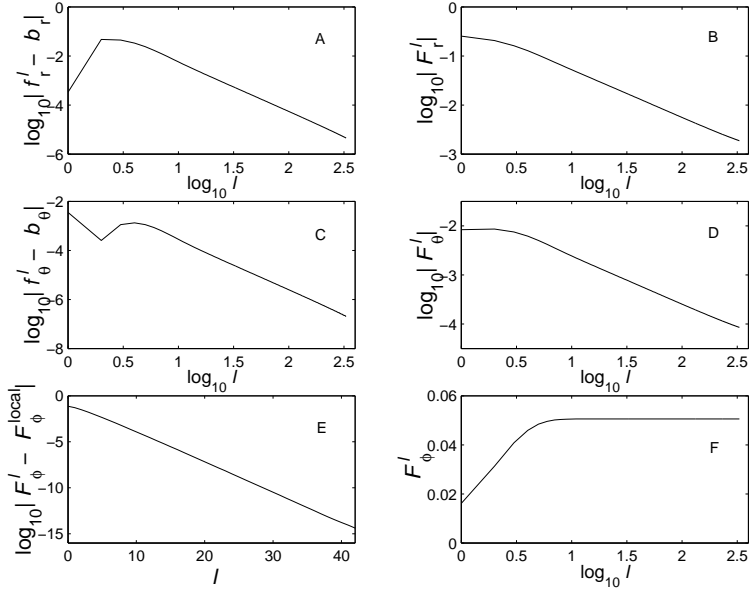,width=10cm,angle=0}
\caption{The regularized SF for a static scalar charge in the
spacetime of a Kerr black hole.
Panels (A) and (C): The $r$ and $\theta$ covariant components of the
individual modes of the regularized self force, respectively,
(i.e., $f^{l}_{r}-b_{r}$ and $f^{l}_{\theta}-b_{\theta}$) as functions
of the mode number $l$.
Panels (B) and (D): $F^{l}_{r}=\sum_{j=0}^{l}(f^{j}_{r}-b_{r})$ and
$F^{l}_{\theta}=\sum_{j=0}^{l}(f^{j}_{\theta}-b_{\theta})$, respectively,
as functions of $l$.
Panel (E): The difference between
$F^{l}_{\varphi}=\sum_{j=0}^{l}(f^{j}_{\varphi}-b_{\varphi})$ and the
local SF, $F^{\rm local}_{\varphi}$, which is given by Eq.\ 
(\ref{kerr:ald}), as a function of $l$.
Panel (F): $F^{l}_{\varphi}$ as a function of $l$. The (unknown) values of
$b_{r}$ and $b_{\theta}$ were approximated by their respective values at
$L=1000$, i.e., by $f^{L=1000}_{r}$ and $f^{L=1000}_{\theta}$, 
correspondingly. The parameters for the data presented here are:
$r=2.2M$, $\theta=\pi/4$, $a=0.2M$, $Q=0$, and $M=1$.}
\label{forcekerr}
\end{figure}

Figure \ref{forcekerr} shows the behavior of the regularized SF. We choose
here $l=330$ and $L=1000$. In Fig.\ 
\ref{forcekerr}(A) and \ref{forcekerr}(B) we show $f_r^l-b_r$ and $F^l_r$
as functions of $l$, respectively, and in Fig.\
\ref{forcekerr}(C) and \ref{forcekerr}(D) we show
$f_{\theta}^l-b_{\theta}$ and $F^l_{\theta}$ as functions of $l$,
respectively. As expected, we find both $f_r^l-b_r$ [Fig.\
\ref{forcekerr}(A)] and
$f_{\theta}^l-b_{\theta}$ [Fig.\ \ref{forcekerr}(C)] to behave like
$l^{-2}$ for large values of $l$. We also find that 
$F^l_r$ [Fig.\ \ref{forcekerr}(B)] and $F^l_{\theta}$ [Fig.\
\ref{forcekerr}(D)] scale like $l^{-1}$ for large values of $l$. When
extrapolated, we conclude that as $l\to\infty$, $F^l_r\to F_r=0$ and 
$F^l_{\theta}\to F_{\theta}=0$. 

\begin{figure}
\epsfig{file=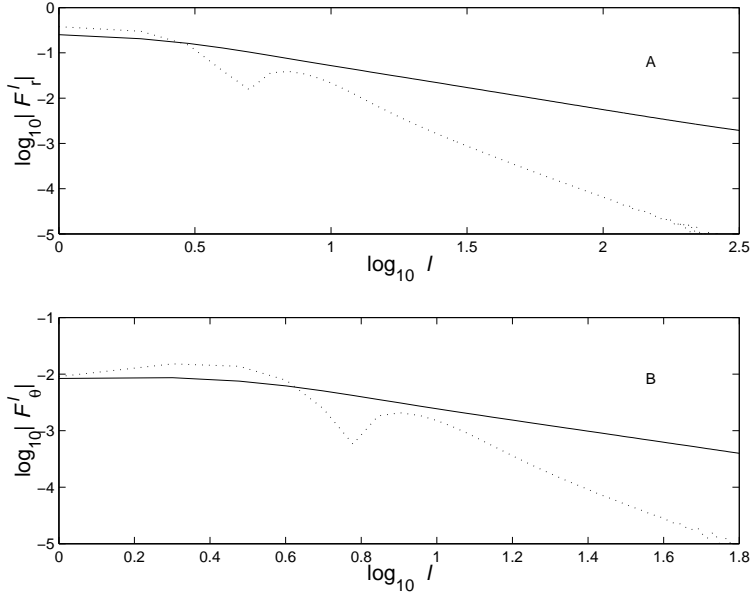,width=10cm,angle=0}
\caption{The regularized SF for a static scalar charge in the spacetime
of Kerr. 
Upper panel (A): $F^{l}_{r}$ without the inclusion of ${\cal
R}^{l+1}_{r}$ and ${\cal E}^{L}_{r}$ [same as in Fig.\ \ref{forcekerr}(B)] 
(solid line) and with their inclusion 
(dotted line). 
Lower panel (B): $F^{l}_{\theta}$ without the inclusion of ${\cal
R}^{l+1}_{\theta}$ and ${\cal E}^{L}_{\theta}$ 
[same as in Fig.\ \ref{forcekerr}(D)] 
(solid line) and with their
inclusion (dotted line).}
\label{erkerr}
\end{figure}

We can approximate $F^l_{r}$ and $F^l_{\theta}$ even better by including 
approximate values for ${\cal R}^{l+1}_{\mu}$ and ${\cal E}^{L}_{\mu}$.
This is done in Fig.\ 
\ref{erkerr}. In Fig.\ \ref{erkerr}(A) we show $F^l_{r}$ twice: without
the inclusion of ${\cal R}^{l+1}_{r}$ and ${\cal E}^{L}_{r}$ [same as in
Fig.\ \ref{forcekerr}(B)] and with their inclusion. We find that for large
values of $l$, the latter behaves like $l^{-2}$ (the former scales like
$l^{-1}$). Similarly, 
in Fig.\ \ref{erkerr}(B) we show $F^l_{\theta}$ twice: without
the inclusion of ${\cal R}^{l+1}_{\theta}$ and ${\cal E}^{L}_{\theta}$
[same as in Fig.\ \ref{forcekerr}(D)] and with their inclusion. Again, 
we find that for large values of $l$, the latter behaves like $l^{-2}$ 
(the
former scales like $l^{-1}$). Recall that we are using here only
approximated values for ${\cal R}^{l+1}_{\mu}$ and ${\cal E}^{L}_{\mu}$. 
For this reason, we do not expect their inclusion to yield an exact zero
result for the $r$ and $\theta$ components of the SF. However, they do
eliminate the leading order term in $F^l_{r}$ and $F^l_{\theta}$, such
that instead of a leading $l^{-1}$ behavior we find an $l^{-2}$
behavior. When this behavior is extrapolated to $l\to\infty$, we again
infer that $F^l_{r}\to F_{r}=0$ and $F^l_{\theta}\to F_{\theta}=0$.

In Fig.\ \ref{forcekerr}(F) we show $F_{\varphi}^l$ as a function of $l$,
and in Fig.\ \ref{forcekerr}(E) we show the difference between
$F_{\varphi}^l$ and $F_{\varphi}^{\rm local}$, which 
is given by Eq.\ (\ref{kerr:ald}).  We find that  
$F_{\varphi}^l$ approaches a non-zero value as $l\to\infty$. We also find
that the difference between $F_{\varphi}^l$ and $F_{\varphi}^{\rm local}$
decays exponentially in $l$ for large-$l$ values. We infer that as
$l\to\infty$, $F_{\varphi}^l\to F_{\varphi}=F_{\varphi}^{\rm local}$. 
We find similar results also for other choices of the parameters. 

When our results for the SF are combined, we find that the tail part of
the regularized SF vanishes. That is, the tail part is given by 
\begin{equation}
F^{\rm tail}_{\mu}=q^2\int_{-\infty}^{\tau -}\,d\tau \nabla_{\mu}
G^{\rm tail}=\sum_{j=0}^{\infty}(f_{\mu}^j-b_{\mu})-d_{\mu}\, .
\end{equation}
According to the conjecture (\ref{d-conjecture}), $d_{\mu}=F_{\mu}^{\rm
local}$. We find numerically that
$\sum_{j=0}^{\infty}(f_{\mu}^j-b_{\mu})=F_{\mu}^{\rm local}$.
Consequently, we infer that $F^{\rm tail}_{\mu}=0$.

\subsection{Kerr-Newman}\label{subsec:kn}
Next, we endow the black hole with an electric charge $Q$ in addition to 
its
spin parameter $a$. In this case, too, we do not have the analytical
results for the MSRP parameters $a_{\mu},b_{\mu}$ and $c_{\mu}$. As in the
Kerr case above, these parameters can be obtained from the large-$l$
behavior of the modes of the bare SF. Again, we shall use the conjecture
(\ref{d-conjecture}) 
regarding the MSRP parameter $d_{\mu}$, according to which  $d_{\mu}$
equals the sum of the two local terms in the full, regularized SF. As the
Kerr-Newman spacetime has non-vanishing Ricci curvature, both terms
contribute. We find that for a static scalar charge outside a  Kerr-Newman
black hole 
\begin{equation}
F_{\mu}^{\rm ALD}=
\frac{1}{3}q^2a\Delta\sin^2\theta\frac{M^2\Sigma+Q^2(Q^2-2Mr)}
{(\Delta-a^2\sin^2\theta)^{5/2}\Sigma^{3/2}}\, \delta^{\varphi}_{\mu}\, ,
\end{equation}
and
\begin{equation}
F_{\mu}^{\rm Ricci}=-
\frac{1}{3}q^2a\Delta\sin^2\theta\frac{Q^2}
{(\Delta-a^2\sin^2\theta)^{3/2}\Sigma^{3/2}}\, \delta^{\varphi}_{\mu}\, ,
\end{equation}      
such that the total local piece of the SF is given by
\begin{equation}
F_{\mu}^{\rm local}=
\frac{1}{3}q^2a\Delta\sin^2\theta\frac{M^2-Q^2}
{(\Delta-a^2\sin^2\theta)^{5/2}\Sigma^{1/2}}\, \delta^{\varphi}_{\mu}\, .
\label{f-local-kn}
\end{equation}      

\begin{figure}
\epsfig{file=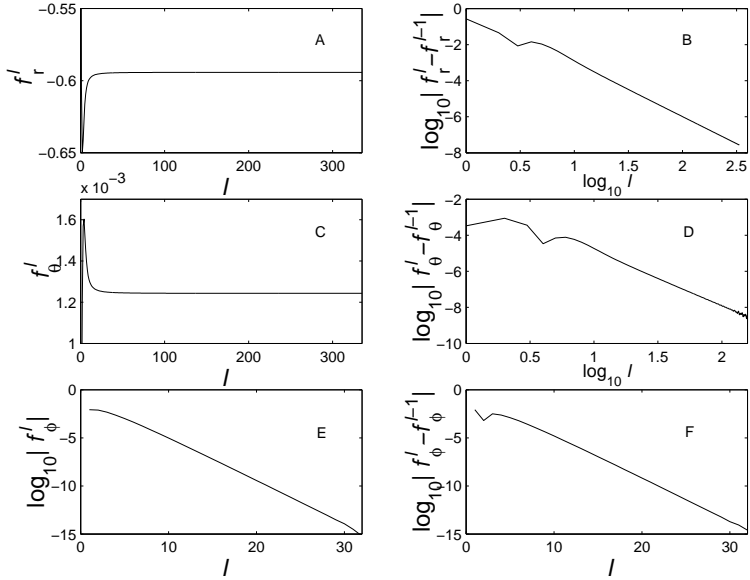,width=10cm,angle=0}
\caption{Behavior of the individual modes of the self force for a static 
scalar charge in the spacetime of a Kerr-Newman black hole.  
Panels (A),
(C), and (E): The $r$, $\theta$, and $\varphi$ covariant components of the 
individual modes of the self force, respectively 
(i.e., $f^{l}_{r}$, $f^{l}_{\theta}$, and $f^{l}_{\varphi}$), as functions
of the mode number $l$. Panels (B), (D), and (F): The $r$, $\theta$,
and $\varphi$ covariant components of the difference between
two consecutive modes of the self force, respectively
(i.e., $f^{l}_{r}-f^{l-1}_{r}$, $f^{l}_{\theta}-f^{l-1}_{\theta}$, and
$f^{l}_{\varphi}-f^{l-1}_{\varphi}$), as functions
of the mode number $l$. The parameters for the data presented here are: 
$r=2.2M$, $\theta=\pi/4$, $a=0.1M$, $Q=0.1M$, and $M=1$.}
\label{modeskn}
\end{figure}

First, we study the behavior of the individual modes
$f^{l}_{\mu}$ of the bare SF, which we compute numerically 
from Eqs.\ (\ref{fr-formal})--(\ref{fphi-formal}). Figure \ref{modeskn} shows
the individual
modes, $f^{l}_{\mu}$, and the
difference between two consecutive modes, $f^{l}_{\mu}-f^{l-1}_{\mu}$, as
functions of the mode number $l$ for $\mu=r, \theta, \varphi$. In Fig.\
\ref{modeskn} we present the first 336 modes (i.e., $l=0,\dots,335$) for
the parameters $r=2.2M$, $\theta=\pi/4$, $a=0.1M$, $Q=0.1M$, and $M=1$,
but similar behavior was found also for other choices for the values of
the parameters. We find that the $r$ and $\theta$ components of the modes
approach a non-zero limiting value in the large-$l$ limit, as the
difference between two consecutive modes behaves like $l^{-3}$ for large
values of $l$. This behavior of the individual modes implies that 
the MSRP parameters $a_r$, $a_{\theta}$, $c_r$, and $c_{\theta}$
vanish. (Recall that we are using here the ``averaged'' value for $a_r$
and $a_{\theta}$. The ``one-sided'' values are, in general,
non-zero.) (Note, that a non-zero value for $c_r$ or $c_{\theta}$ implies
that the difference between two consecutive modes should scale like
$l^{-2}$.) However, the MSRP parameters $b_r$ and $b_{\theta}$ are
non-zero. These parameters correspond to the $l\to\infty$ limit of the 
individual modes $f^{l}_{r}$ and $f^{l}_{\theta}$, respectively (recall
that $a_r$ and $a_{\theta}$ vanish). The $\varphi$ component of the
individual modes of the SF drops off exponentially for large values of
$l$, and we infer that $a_{\varphi}=0$, $b_{\varphi}=0$, and  
$c_{\varphi}=0$. This is in agreement with the weak-field expansion 
of Section \ref{weak}.

\begin{figure}
\epsfig{file=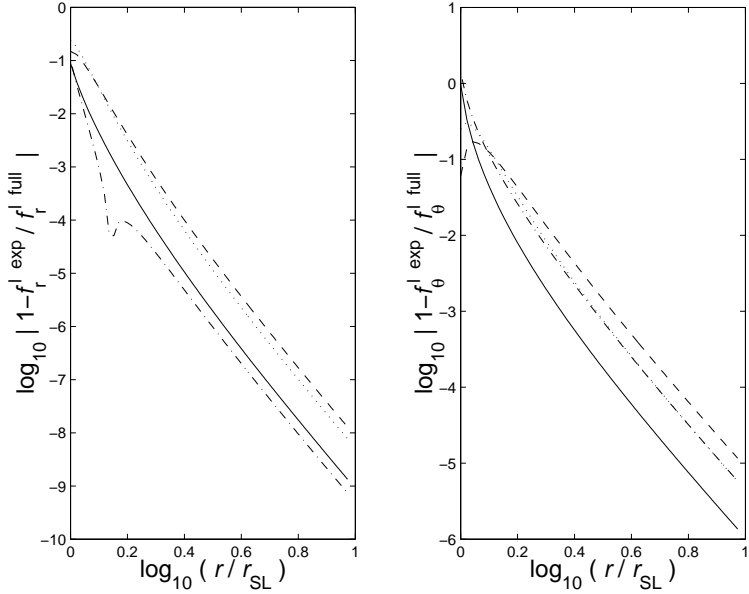,width=10cm,angle=0}
\caption{
Comparison of the modes of the bare force as given by the asymptotic
expansions [Eq.\ (\ref{flr1}) for the $r$ component and Eq.\
(\ref{flth1}) for the $\theta$ component] and the full expressions which we
compute numerically from Eqs.\ (\ref{fr-formal}) and 
(\ref{fth-formal}), respectively. 
We show the relative difference of the two
expressions as a function of $r/r_{\rm SL}$ for a number of mode numbers
$l$. 
Left panel: the $r$ component. Right panel: the $\theta$  component. In
both panels we show the modes: $l=1$ (solid line), $l=2$ (dash-dotted
line), $l=3$ (dashed line), and $l=4$ (dotted line). The data here are
shown for the following values for the parameters: 
$M=1$, $a=0.6M$, $Q=0.4M$, and $\theta=\pi/4$.}
\label{comparison}
\end{figure}

Next we compare our results for the modes here, with our asymptotic
expansions in Section \ref{weak}. Figure \ref{comparison} displays the
relative difference of the asymptotic values for the modes of the bare
SF [given by Eq.\ (\ref{flr1}) for the $r$ component and by Eq.\ 
(\ref{flth1}) for the $\theta$ component] 
and the full values of the modes of the bare SF [which we compute
numerically from Eqs.\ (\ref{fr-formal}) and 
(\ref{fth-formal})] as functions of $r$ (in units of
$r_{\rm SL}=M+\sqrt{M^2-a^2 \cos^2\theta}$, the $r$ value at the static 
limit) for different values of
the mode number $l$. At large distances we
find both the $r$ and $\theta$ components to agree with the asymptotic
expansion at the right orders. Near the static limit, they of course
disagree. Note, however, that even very close to the static limit the $r$
component of the modes 
of the asymptotic expansion does not deviate from their full-expression 
counterparts by much more than $10\%$. We find similar results also for
other choices of the parameters. 

\begin{figure}
\epsfig{file=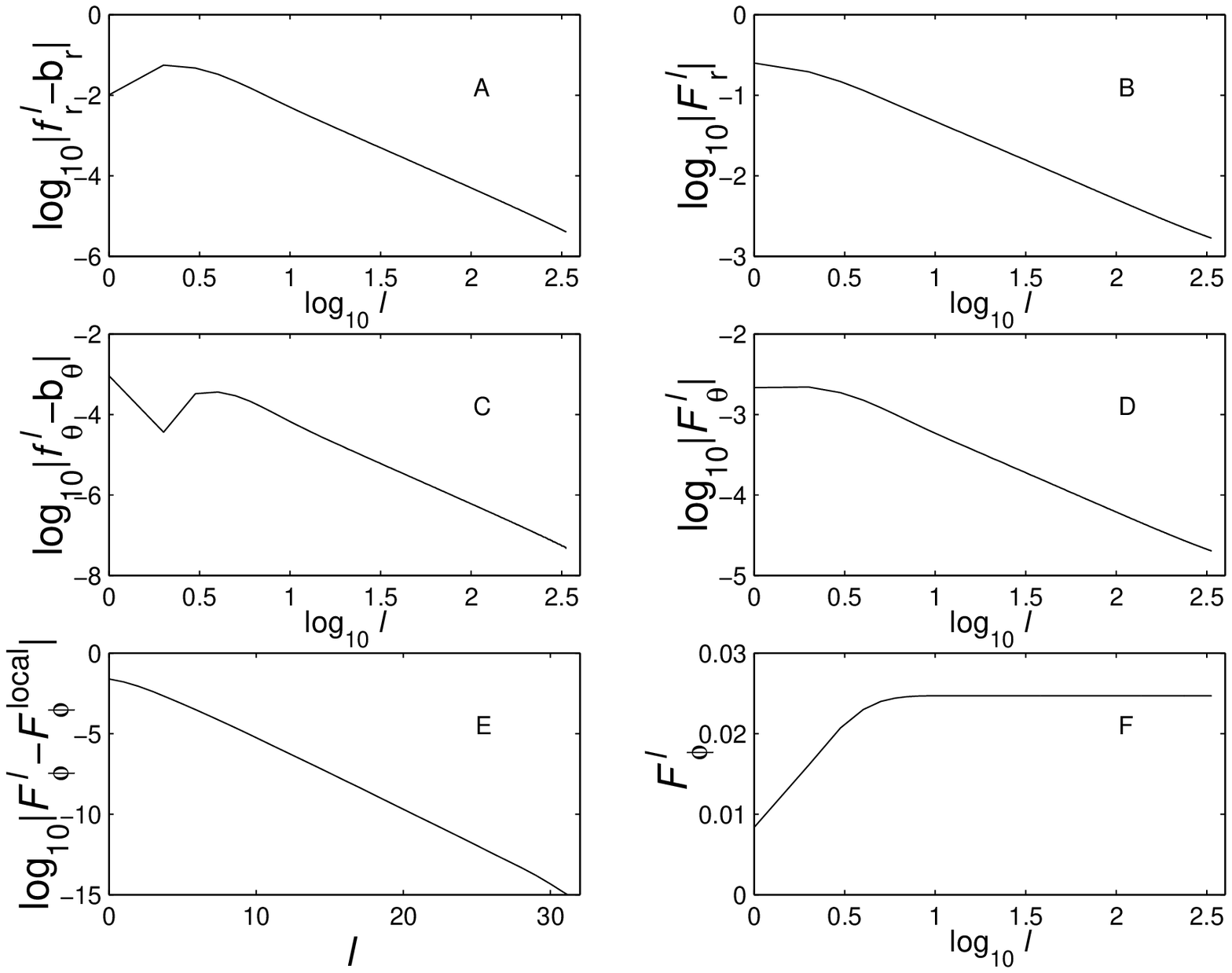,width=10cm,angle=0}
\caption{The regularized SF for a static scalar charge in the
spacetime of a Kerr-Newman black hole.
Panels (A) and (C): The $r$ and $\theta$ covariant components of the
individual modes of the regularized self force, respectively,
(i.e., $f^{l}_{r}-b_{r}$ and $f^{l}_{\theta}-b_{\theta}$) as functions
of the mode number $l$.
Panels (B) and (D): $F^{l}_{r}=\sum_{j=0}^{l}(f^{j}_{r}-b_{r})$ and
$F^{l}_{\theta}=\sum_{j=0}^{l}(f^{j}_{\theta}-b_{\theta})$, respectively,
as functions of $l$.
Panel (E): The difference between
$F^{l}_{\varphi}=\sum_{j=0}^{l}(f^{j}_{\varphi}-b_{\varphi})$ and the
local SF, $F^{\rm local}_{\varphi}$, as a function of $l$.
Panel (F): $F^{l}_{\varphi}$ as a function of $l$. The (unknown) values of
$b_{r}$ and $b_{\theta}$ were approximated by their respective values at
$L=1000$, i.e., by $f^{L=1000}_{r}$ and $f^{L=1000}_{\theta}$,
correspondingly. 
The parameters for the data presented here are: 
$r=2.2M$, $\theta=\pi/4$, $a=0.1M$, $Q=0.1M$, and $M=1$.}
\label{forcekn}
\end{figure}

Figure \ref{forcekn} displays the behavior of the regularized SF. We
choose here again $l=330$ and $L=1000$. 
In Fig.\
\ref{forcekn}(A) and \ref{forcekn}(B) we show $f_r^l-b_r$ and $F^l_r$
as functions of $l$, respectively, and in In Fig.\
\ref{forcekn}(C) and \ref{forcekn}(D) we show
$f_{\theta}^l-b_{\theta}$ and $F^l_{\theta}$ as functions of $l$,
respectively. Similar to the Kerr case, we find both $f_r^l-b_r$ [Fig.\
\ref{forcekn}(A)] and $f_{\theta}^l-b_{\theta}$ [Fig.\
\ref{forcekn}(C)] to behave like $l^{-2}$ for large values of $l$. We
also find that $F^l_r$ [Fig.\ \ref{forcekn}(B)] and $F^l_{\theta}$
[Fig.\ \ref{forcekn}(D)] scale like $l^{-1}$ for large values of
$l$. When extrapolated, we conclude that as $l\to\infty$, $F^l_r\to F_r=0$
and $F^l_{\theta}\to F_{\theta}=0$.            

\begin{figure}
\epsfig{file=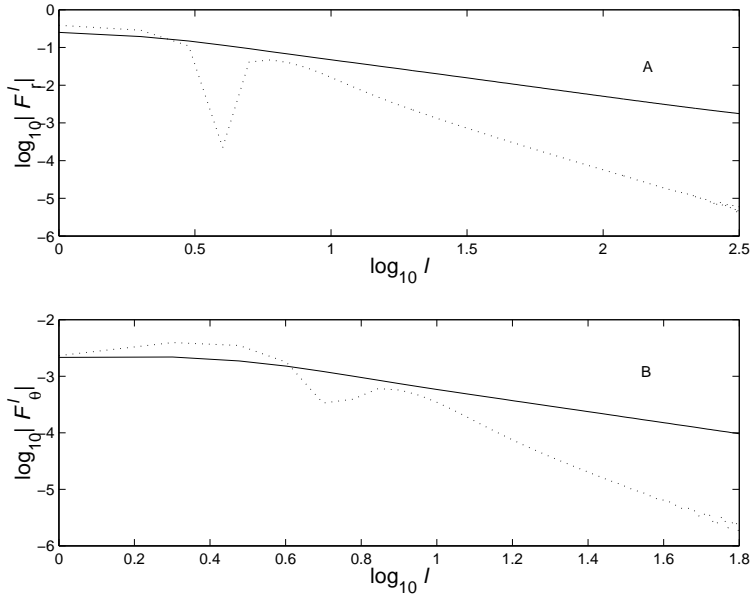,width=10cm,angle=0}
\caption{The regularized SF for a static scalar charge in the spacetime
of a Kerr-Newman black hole. 
Upper panel (A): $F^{l}_{r}$ without the inclusion of ${\cal
R}^{l+1}_{r}$ and ${\cal E}^{L}_{r}$ [same as in Fig.\ \ref{forcekn}(B)]
(solid line) and with their inclusion
(dotted line).
Lower panel (B): $F^{l}_{\theta}$ without the inclusion of ${\cal
R}^{l+1}_{\theta}$ and ${\cal E}^{L}_{\theta}$
[same as in Fig.\ \ref{forcekn}(D)]
(solid line) and with their
inclusion (dotted line).}
\label{erkn}
\end{figure}

In a similar way to our analysis of the case of a Kerr spacetime, 
we can approximate $F^l_{r}$ and $F^l_{\theta}$ even better by including
${\cal R}^{l+1}_{\mu}$ and ${\cal E}^{L}_{\mu}$. This is done in Fig.\
\ref{erkn}. In Fig.\ \ref{erkn}(A) we show $F^l_{r}$ twice: without
the inclusion of ${\cal R}^{l+1}_{r}$ and ${\cal E}^{L}_{r}$ [same as in
Fig.\ \ref{forcekn}(B)] and with their inclusion. We again find that for
large values of $l$, the latter behaves like $l^{-2}$ (the former scales
like $l^{-1}$). Similarly,
in Fig.\ \ref{erkn}(B) we show $F^l_{\theta}$ twice: without
the inclusion of ${\cal R}^{l+1}_{\theta}$ and ${\cal E}^{L}_{\theta}$
[same as in Fig.\ \ref{forcekn}(D)] and with their inclusion. Again,
we find that for large values of $l$, the latter behaves like $l^{-2}$
(the
former scales like $l^{-1}$). Recall that we are using here only
approximated values for ${\cal R}^{l+1}_{\mu}$ and ${\cal E}^{L}_{\mu}$.
For this reason, we do not expect their inclusion to yield an exact zero
result for the $r$ and $\theta$ components of the SF. However, like in 
the Kerr case, they do eliminate the leading order term in $F^l_{r}$ and
$F^l_{\theta}$, such that instead of a leading $l^{-1}$ behavior we find
an $l^{-2}$ behavior. When this behavior is extrapolated to $l\to\infty$,
we again infer that $F^l_{r}\to F_{r}=0$ and 
$F^l_{\theta}\to F_{\theta}=0$.               

In Fig.\ \ref{forcekn}(F) we show $F_{\varphi}^l$ as a function of $l$,
and in Fig.\ \ref{forcekn}(E) we show the difference between
$F_{\varphi}^l$ and $F_{\varphi}^{\rm local}$, which is given 
by Eq.\ (\ref{f-local-kn}).  We find that
$F_{\varphi}^l$ approaches a non-zero value as $l\to\infty$. We also find
that the difference between $F_{\varphi}^l$ and $F_{\varphi}^{\rm local}$
decays exponentially in $l$ for large-$l$ values. We infer that as
$l\to\infty$, $F_{\varphi}^l\to F_{\varphi}=F_{\varphi}^{\rm local}$.   
We find similar results also for other choices of the parameters.

As in the case of a Kerr spacetime, when our results for the SF are
combined, we find that the tail part of the regularized SF vanishes. 

\section{Far field computation of the self force}\label{sec:farfield}

In Sections \ref{weak} and \ref{sec:num} we computed the SF  
by using the field (and its gradient) evaluated on the particle's 
world line, i.e., by using the near field. In this Section, we shall 
compute the SF using the far field (evaluated asymptotically at infinity 
and at the black hole's event horizon), and demonstrate the compatibility 
of the two approaches. For simplicity, we shall restrict our considerations 
in this Section to the case of a Kerr black hole. Specifically, 
we shall show that the covariant $t$ and 
$\varphi$ components of the SF ($F_t$ and $F_{\varphi}$, respectively) 
in the case of a Kerr black hole can also be inferred from balance arguments 
pertaining to the global conservation of energy and angular momentum. 
Specifically, we shall deduce 
$F_t$ and $F_{\varphi}$ by calculating the fluxes of energy and angular 
momentum, associated with the charge's scalar field, 
flowing out to infinity and down the black hole's event horizon. 
We shall show that the results of these far-field calculations agree with the
near-field calculations we have performed in Sections \ref{weak} and 
\ref{sec:num}. In fact, by showing this agreement we demonstrate the applicability 
of the MSRP, and the validity of the conjecture (\ref{d-conjecture}) to the 
problem of interest. 

Since we need to evaluate the scalar field near the horizon and at
infinity, we place the particle at $(r_0,\theta_0,\varphi_0)$ in 
this Section to avoid confusion.

The rate of change of the particle's four-momentum due to the SF,  
$F_{\mu}$, is given by 
\begin{equation}
  \frac{dp_{\mu}}{d\tau}=F_{\mu} \ ,
\end{equation}
where $\tau$ is the proper time. In particular, we have
\begin{equation}
  F_{\varphi}=\frac{dp_{\varphi}}{d\tau}=
\frac{1}{\sqrt{-g_{tt}}}\, \frac{dL}{dt} \ ,
\label{fphi-balance}
\end{equation}
where $L\equiv p_{\varphi}$ is the angular momentum of the particle 
along the black hole's rotation axis, and we have used the fact that 
$dt/d\tau=1/\sqrt{-g_{tt}}$ (recall that the charge is static).
The sum of the rate of change of the particle's angular momentum 
(i.e., $\,dL/\,dt$), 
the total amount of angular momentum
(per unit time) flowing out to infinity (${\mathcal F}_{\infty}^L$) and
down the black hole's event horizon (${\mathcal F}_{\rm hole}^L$) must be zero, 
if global angular momentum were to be conserved. Hence, 
\begin{equation}
-\,dL/\,dt={\mathcal F}_{\infty}^L+{\mathcal F}_{\rm hole}^L \ .
\end{equation}
The value of 
${\mathcal F}_{\infty}^L$ is given by (see Chapter~5 of Ref.~\cite{MTW}) 
\begin{equation}
  {\mathcal F}_{\infty}^L=\lim_{r \rightarrow \infty} \int 
T_{\mu}^{\ r}\, \xi_{(\varphi)}^{\mu}\, r^2\, d\Omega \label{flux-inf} \ ,
\end{equation}
where ${\boldmath{\xi_{(\varphi)}}}=\partial/\partial \varphi$ is the axial 
Killing vector, $d\Omega =\sin \theta\, d\theta\, d\varphi$, and the 
stress-energy tensor $T_{\mu \nu}$ associated with the scalar field $\Phi$ 
is given by
\begin{equation}
  T_{\mu \nu}=\frac{1}{4\pi} \left( \Phi_{,\mu} \Phi_{,\nu}-{1\over 2}g_{\mu \nu}
g^{\alpha \beta} \Phi_{,\alpha} \Phi_{,\beta} \right) \ . \label{Tab}
\end{equation}
The total angular momentum flowing down the event horizon per unit time,  
${\mathcal F}_{\rm hole}^L$, is given by~\cite{hawking-hartle72,teuk-press}
\begin{equation}
  {\mathcal F}_{\rm hole}^L = \lim_{r \rightarrow r_+}\int [-2Mr_+ T_{\mu \nu}\,
\xi_{(\varphi)}^{\mu}\, l^{\nu}_{HH}] \, d\Omega \ , \label{flux-hole}
\end{equation}
where $l^{\mu}_{HH}$ is one of the basis vectors of the
Hawking-Hartle tetrad. It is an outgoing tetrad which is made well behaved 
on the future event horizon~\cite{teuk-73,teuk-press,hawking-hartle72}.
The components of $l^{\nu}_{HH}$, in Boyer-Lindquist coordinates 
$(t,r,\theta,\varphi)$, are given by
\begin{equation}
  l^{\mu}_{HH}=\left[ \frac{1}{2},\frac{\Delta}{2(r^2+a^2)},0,\frac{a}{2(r^2+a^2)}
\right] \ .
\end{equation}
Note that although some of the components of $T_{\mu \nu}$, in Boyer-Lindquist 
coordinates, diverge on the event horizon, ${\mathcal F}_{\rm hole}^L$ remains 
finite. In fact, the divergence is due entirely to the coordinate singularity.
However, ${\mathcal F}_{\rm hole}^L$, being a scalar, is independent of the choice 
of coordinates.

The asymptotic expressions of the scalar field $\Phi$ at infinity and on the horizon
can be deduced from Eqs.~(\ref{phi-mode}), (\ref{sol1}), (\ref{1sol}), (\ref{2sol}),
(\ref{wronz}), (\ref{Klmu}), (\ref{Sr}), (\ref{z}) and~(\ref{Plmu-asy}).
The results are
\begin{eqnarray}
  \Phi(r,\theta,\varphi) &\rightarrow& \sqrt{1-\frac{2Mr_0}{\Sigma_0}}\ \frac{q}{r}
\ \ \ \ \ r \rightarrow \infty \label{Phi-inf} \\
  \Phi(r,\theta,\varphi) &\rightarrow& \sum_{l=0}^{\infty} \sum_{m=-l}^l Z_{lm}\,
e^{-ik_m r_*} Y_{lm}(\theta,\varphi) \ \ \ \ \ r \rightarrow r_+ \ ,
\label{Phi-hole}
\end{eqnarray}
where $r_*$ is defined by $dr_*/dr =(r^2+a^2)/\Delta$, $k_m=-m\omega_+ \equiv
-ma/(2Mr_+)$, and $Z_{lm}$ is given by
\begin{equation}
  Z_{lm}= C_{lm}\,\frac{\sqrt{M^2-a^2}\, S(r_0)\, (1-z_0^2)}{z_0^{l+1}\, (2l+1)!!}
\left[ \prod_{j=1}^l (j+im\gamma)\right] K_l^{\mu}(z_0)\, Y^*_{lm}(\theta_0,\phi_0)
\ .
\end{equation}
Here $C_{lm}$ are complex constants of unit modulus, i.e.\ $|C_{lm}|=1$.
Substituting Eq.~(\ref{Phi-inf}) into Eqs.~(\ref{Tab}) and (\ref{flux-inf}),
we find that ${\mathcal F}_{\infty}^L=0$. This is expected since
the particle is static relative to static observers at infinity, so no
radiation is emitted. The rate of change of the particle's angular 
momentum is then given by
\begin{equation}
  \frac{\,dL}{\,dt}=-{\mathcal F}_{\rm hole}^L = \lim_{r\rightarrow r_+} 
\left[ 2Mr_+ \int T_{\varphi \nu}\, l_{HH}^{\nu}\, d\Omega\, \right] \ .
\end{equation}
Straightforward calculations yield 
\begin{eqnarray}
  {\mathcal F}_{\rm hole}^L &=& 
-\sum_{l=0}^{\infty} \sum_{m=-l}^l \frac{m^2\, a}{4\pi}
|Z_{lm}|^2 \\
  &=& - \sum_{l=0}^{\infty} \sum_{m=-l}^l \frac{a(M^2-a^2)\, S^2(r_0)}
{4\pi\, z_0^{2l-2}} \left( 1-\frac{1}{z_0^2}\right)^2 m^2 \left[
\prod_{j=1}^l (m^2\gamma^2+j^2)\right] \left[ \frac{K_l^{\mu}(z_0)}{(2l+1)!!}
\right]^2 Y^*_{lm}(\theta_0,\varphi_0) Y_{lm}(\theta_0,\varphi_0) \ .
\end{eqnarray}
The covariant $\varphi$-component of the SF, $F_{\varphi}$, is then
calculated by Eq.~(\ref{fphi-balance}). Using Eqs.~(\ref{Sr}) and (\ref{z}),
we finally obtain 
\begin{equation}
  F_{\varphi}=-\sum_{l=0}^{\infty}\sum_{m=-l}^l
\frac{q a S(r_0)}{z_0^{2l}} \left(1-{1\over z_0^2}\right)
m^2 \left[ \prod_{j=1}^l (m^2\gamma^2+j^2)\right]
\left[ \frac{K_l^{\mu}(z_0)}{(2l+1)!!}\right]^2 Y^*_{lm}(\theta_0,\varphi_0)
Y_{lm}(\theta_0,\varphi_0) \ ,
\end{equation}
which agrees with Eq.~(\ref{fphi}) (Recall that the MSRP parameters
$a_{\varphi}=b_{\varphi}=c_{\varphi}=0$).

Similarly, we can deduce the covariant time component of the SF, $F_t$,  
by the energy balance argument. We have 
\begin{equation}
  F_t=\frac{dp_t}{d\tau}=-\frac{1}{\sqrt{-g_{tt}}}\, \frac{dE}{dt} \ ,
\end{equation}
where $E\equiv -p_t$ is the energy of the particle. The sum of the 
rate of change of the particle's energy (i.e., $\,dE/\,dt$), the  
total amount of the energy (per unit time) associated with the scalar field flowing 
out to infinity (${\mathcal F}_{\infty}^E$) and down the horizon 
(${\mathcal F}_{\rm hole}^E$) vanishes because of global conservation of energy, 
i.e., 
\begin{equation}
-\,dE/\,dt={\mathcal F}_{\infty}^E+{\mathcal F}_{\rm hole}^E \ .
\end{equation}
The relevant formulae are~\cite{MTW,hawking-hartle72,teuk-press}:
\begin{eqnarray}
  {\mathcal F}_{\infty}^E &=& \lim_{r\rightarrow \infty} \int -T_{\mu}^{\ r}\,
\xi_{(t)}^{\mu}\, r^2\, d\Omega  \\
{\mathcal F}_{\rm hole}^E &=& \lim_{r\rightarrow r_+} \int 2Mr_+\,
T_{\mu \nu}\, \xi_{(t)}^{\mu}\, l_{HH}^{\nu}\, d\Omega \ ,
\end{eqnarray}
where ${\boldmath{\xi_{(t)}}}=\partial/\partial t$ is a Killing vector.
Straightforward calculations yield  
${\mathcal F}_{\infty}^E=0={\mathcal F}_{\rm hole}^E$. 
Hence $\,dE/\,dt=0=F_t$, as expected. 

\section{Properties of the self force}\label{sf:prop}

From our numerical study we infer that the full, regularized SF on a
static scalar charge in Kerr-Newman is given by 
\begin{equation}
F_{\mu}^{\rm SF}=
\frac{1}{3}q^2a\Delta\sin^2\theta\frac{M^2-Q^2}
{(\Delta-a^2\sin^2\theta)^{5/2}\Sigma^{1/2}}\, \delta^{\varphi}_{\mu}\, .
\label{sf}
\end{equation}        
We note the following properties of this result. 
\begin{itemize}
\item
The SF vanishes as $a\to 0$, as expected. In that limit the
black hole becomes Reissner-Nordstr\"{o}m (or Schwarzschild in the lack of electric 
charge), for which the SF vanishes, as we found in Section \ref{subsec:rn}.
\item When $t\to -t$ the SF reverses its sign. [Notice that the SF has the form of 
$a\times ({\rm even\, powers\, of\, } a)$.] This is indeed expected, because 
under time reversal the black hole reverses its spin, and rotates in the 
opposite direction. This change of sign under time reversal implies that 
this SF is dissipative.  
\item The SF diverges as $\Delta-a^2\sin^2\theta\to 0$. 
This is not surprising, as
$\Delta-a^2\sin^2\theta= 0$ defines the static limit, beyond which (inside
the ergosphere) no timelike static trajectories exist. Consequently, our
problem of finding the SF on a static particle becomes ill posed beyond the
static limit. 
\item The SF is only in the $\partial /\,\partial\varphi$ direction. The
addition to the regular force which is needed to keep the particle in its
static position is, therefore, orthogonal to the regular force (which has
components only in the $\partial /\,\partial r$ and
$\partial /\,\partial\theta$ directions). 
\item Denoting the
magnitude of the acceleration of the static particle in the absence of the
SF by $a^{\rm reg}$, and the magnitude of the acceleration due only to the
SF by $a^{\rm SF}$, we find the ratio of the two accelerations to be given
by
\begin{equation}
\frac{a^{\rm SF}}{a^{\rm reg}}=
\frac{1}{3}\frac{q^2}{\mu}a\sin\theta\frac{\Sigma^{1/2}\Delta^{1/2}}
{\Delta-a^2\sin^2\theta}\left\{\Delta\left[M\left(r^2-a^2\cos^2\theta\right)
-rQ^2\right]^2
+\frac{1}{4}a^4\left(2Mr-Q^2\right)^2\sin^22\theta\right\}^{-1/2}\, ,
\label{ratio}
\end{equation}
where $\mu$ is the mass of the scalar charge. Clearly, this ratio 
diverges as the static limit is approached. In this sense, the SF is not a
tiny correction for the regular external force which is exerted in order
to keep the particle fixed: the SF becomes dominant over the regular
acceleration when the particle is 
sufficiently close to the static limit. 
The origin of the divergence of this ratio is in the ``damping part'' of
the ALD force. 
The ``Schott part'' of the ALD force diverges too approaching the static
limit, but at a slower rate than the ``damping part''. 
The part of the SF which couples to Ricci curvature
diverges at the same rate as $a^{\rm reg}$ approaching the static limit.
The leading order divergence in Eq.\ (\ref{ratio}) comes 
then from the ``damping part'' of
the ALD force.  This can be readily understood by the following argument.
The ``damping part'' of the ALD force diverges like ${a^{\rm reg}}^2$, 
such that its ratio to $a^{\rm reg}$ is expected to diverge
like $a^{\rm reg}$. Indeed, $a^{\rm reg}$ diverges like 
$(\Delta-a^2\sin^2\theta)^{-1}$ approaching the static limit. 

The scalar field theory does not restrict the charge-to-mass ratio of a
scalar particle, $q/\mu$. (For other field theories we find the
charge-to-mass ratios to
span many orders of magnitude: it is unity for a gravitational charge, but
$2\times 10^{21}$ for an electron.) We can thus view $q/\mu$ as a free
parameter in Eq.\ (\ref{ratio}). Recall, however, that $q/M$ is
assumed to be a small quantity. As $r\to\infty$ the ratio 
$a^{\rm SF}/a^{\rm reg}\to 0$. Thus, there is a value of $r$ (outside the
static limit) where $a^{\rm SF}=a^{\rm reg}$. (Recall, however, that the
SF acceleration is in the $\partial /\,\partial\varphi$ direction, while
the regular acceleration is in the $\partial /\,\partial r$ and 
$\partial /\,\partial\theta$ directions.)
Specializing now to the equatorial plane in Kerr, we find that $a^{\rm
SF}$ equals $a^{\rm reg}$ at
\begin{eqnarray}
r=\frac{2}{3}M&+&\frac{2^{-\frac{1}{3}}}{3}M^{\frac{2}{3}}\mu^{-\frac{1}{3}}
\left[27a\frac{q^2}{M}+16\mu
M+3\sqrt{3a\frac{q^2}{M}\left(27a\frac{q^2}{M}+32\mu M\right)}\right]
^{\frac{1}{3}}\nonumber \\
&+&\frac{2^{\frac{7}{3}}}{3}\mu^{\frac{1}{3}}
M^{\frac{4}{3}}\left[27a\frac{q^2}{M}+16\mu M+3
\sqrt{3a\frac{q^2}{M}\left(27a\frac{q^2}{M}+32\mu M\right)}\right]
^{-\frac{1}{3}}.
\end{eqnarray}
For small values of $aq^2/(\mu M^2)$, this value of $r$ can be expanded in
powers of $aq^2/(\mu M^2)$. We find that 
$$r=2M\left\{1+\frac{1}{24}\left(\frac{q}{\mu}\right)
\left(\frac{q}{M}\right)\left(\frac{a}{M}\right)
-\frac{1}{288}\left(\frac{q}{\mu}\right)^2
\left(\frac{q}{M}\right)^2\left(\frac{a}{M}\right)^2
+O\left[
\left(\frac{q}{\mu}\right)^3
\left(\frac{q}{M}\right)^3\left(\frac{a}{M}\right)^3\right]\right\}
\, ,$$ such that for small $aq^2/(\mu M^2)$ the two accelerations become
comparable only very close to the static limit. 
Figure \ref{accel} displays this value for $r$ vs.\ the free parameter 
$q/\mu$ for three values of $q/M$. Keeping $a/M$ and $q/M$ fixed, we find
that at small values of $q/\mu$
the two accelerations become comparable only very close to the static
limit. For large values of $q/\mu$ they become equal at values of $r$
which scale like $(q/\mu)^{1/3}$. The change in the behavior occurs near
$q/\mu\approx M^2/(aq)$. Recalling that $q/M\ll 1$ and that $a/M<1$, we
find that at the change of behavior in Fig.\ \ref{accel}, $q/\mu\gg 1$.
In order to have $r\gg M$ (the distance from the black hole at which
the two
accelerations become comparable is very large) we should thus require 
$q/\mu\gg M^2/(aq)$.
\begin{figure}
\epsfig{file=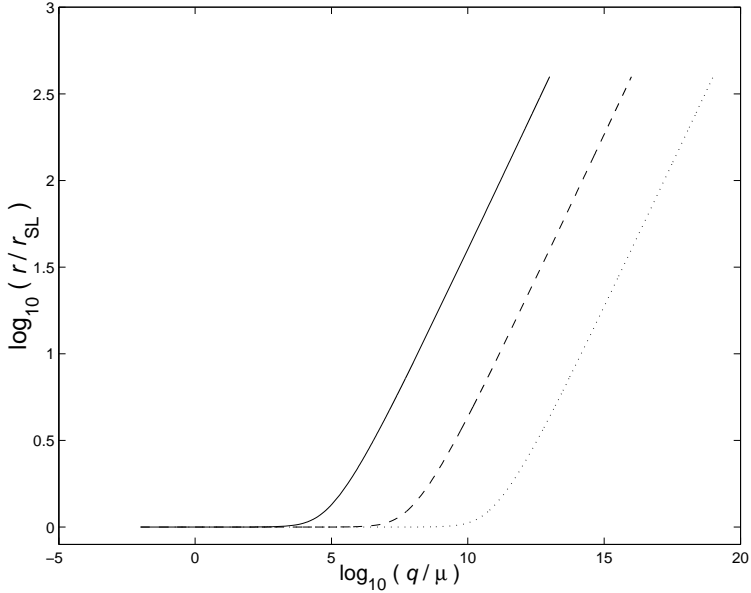,width=10cm,angle=0}
\caption{The value of $r/r_{\rm SL}$ where $a^{\rm SF}$ equals 
$a^{\rm reg}$ as a
function of $q/\mu$ on the equatorial plane of a Kerr black hole for
$a/M=0.5$, for three values of $q/M$: $q/M=10^{-4}$ (solid line), 
$q/M=10^{-7}$ (dashed line), and $q/M=10^{-10}$ (dotted line).
}
\label{accel}
\end{figure}

\item At large distances this SF agrees with the SF found by
Gal'tsov \cite{galtsov-82}, i.e., at large distances we find that 
\begin{equation}
F_{\varphi}=\frac{q^2}{3}a\sin^2\theta\frac{M^2-Q^2}{r^4}\left\{
1+3{M\over r}+\frac{1}{2}[3(5M^2-Q^2)+2a^2(1-3\cos^2\theta)]
\frac{1}{r^2} + O\left(r^{-3}\right)
\right\}\, ,
\end{equation}
whose leading term agrees with Gal'tsov's result when $Q=0$ (a Kerr black
hole). Notice, that this expansion coincides with the asymptotic solution
we found above in Eq.\ (\ref{exp-1}).

\item The direction of the SF is in the direction of the spin of the black
hole. Namely, in order to hold the particle static, the applied external
force should be in the direction opposite to the spin. As noted by
Gal'tsov \cite{galtsov-82}, this direction can be explained as a tidal
friction effect \cite{hawking-hartle72}: As the particle is accelerated in
the direction of the black hole's spin, global conservation of
angular momentum implies that the black hole is accelerated in the
direction opposite to the spin, such that the black hole tends to spin
down, and its rotational energy is being dissipated.

\item When the particle's position is off the black hole's polar axis, the black hole is 
immersed in an external field, such that the entire configuration is not axially 
symmetric. From Hawking's theorem \cite{hawking-72}, stating that a stationary black 
hole must be either static or axisymmetric, it then follows that the black hole cannot 
remain stationary: it must evolve in time until it has become static or until it 
has achieved an axisymmetric orientation with the external field \cite{press-72}. As the 
scalar field is stationary, there can't be any flux of energy down the event horizon, as 
measured by a static observer at infinity. Hence the black hole's mass $M$ is unchanged. 
As the black hole's surface area ${\cal A}$, given by 
${\cal A}=8\pi M(M+\sqrt{M^2-a^2})$ must 
increase, it follows that its angular momentum $a$ must decrease, or that the black 
hole spins down. The dissipated rotational energy of the black hole does not escape to 
infinity, as the field is strictly static there. Instead, it flows down the hole as seen 
by a local, dragged observer. Specifically, a local observer who follows a timelike 
orbit must be dragged inside the ergosphere. Any such observer will see the field as 
time dependent, and will see a flux of energy down the event horizon, whose origin is in 
the black hole's rotational energy. However, when the particle is on the black hole's 
polar axis, the SF vanishes according to Eq.~(\ref{sf}). In this case, 
the black hole is immersed in an axisymmetric field, such that it can 
remain stationary. 
The flux of angular momentum as viewed by a static distant observer 
vanishes, and the black hole's spin is unchanged.

\item When Newton's constant $G$ and the speed of light $c$ are
re-introduced, and the SF (\ref{sf}) is expanded in powers of $G/c^2$, we
find that 
\begin{eqnarray}
F^{\rm SF}_{\varphi}&=&
\frac{q^2}{3}\left(\frac{G}{c^2}\right)^2a\sin^2\theta\frac{M^2-G^{-1}Q^2}
{r^4}\left\{\left[1+2\left(1-3\cos^2\theta\right)\frac{a^2}{r^2}\right]
+\left[3+\left(5-14\cos^2\theta\right)\frac{a^2}{r^2}\right]
\left(\frac{G}{c^2}\right)\frac{M}{r}\right.
\nonumber \\
&+&\left. O\left(\frac{G}{c^2}\right)^2\right\}\, ,
\label{pn}
\end{eqnarray}
such that we find that this SF is a (post)$^2$-Newtonian  
effect. Note, that at the (post)$^2$-Newtonian 
order the effect has contributions both from the $r^{-4}$
and the $r^{-6}$ terms. We also write explicitly in Eq.\ (\ref{pn}) the 
(post)$^3$-Newtonian  order of the SF. 
\end{itemize}

\section*{Acknowledgements}
This research was supported by NSF grants AST-9731698 and PHY-9900776 and
by NASA grant NAG5-6840.            

\begin{appendix}

\section{}\label{app-a}
\begin{center}
{\bf EVALUATION OF {\boldmath $X_l(p,\theta)$} AND {\boldmath $\xi_l(p,\theta)$}}
\end{center}

When calculating the SF in the weak field region, we encounter the functions 
$X_l(p,\theta)$ and $\xi_l(p,\theta)$ defined as
\begin{eqnarray}
  X_l(p,\theta) &=& \sum_{m=-l}^l m^p Y^*_{lm}(\theta,\varphi) 
Y_{lm}(\theta,\varphi) \label{Xl1} \\
  \xi_l(p,\theta) &=& \sum_{m=-l}^l m^p Y^*_{lm}(\theta,\varphi) 
\frac{\partial}{\partial \theta} Y_{lm}(\theta,\varphi) \ .
\end{eqnarray}
The spherical harmonics are of the form 
$Y_{lm}(\theta,\varphi)={\cal Y}_{lm}(\theta) 
e^{im\varphi}$, where ${\cal Y}_{lm}(\theta)$ is a real-valued function. It follows 
that $X_l(p,\theta)$ and $\xi_l(p,\theta)$ are independent of $\varphi$. 
So we can take $\varphi=0$ and write
\begin{eqnarray}
  X_l(p,\theta) &=& \sum_{m=-l}^l m^p [Y_{lm}(\theta,0)]^2 \\
  \xi_l(p,\theta) &=& \sum_{m=-l}^l m^p Y_{lm}(\theta,0) \frac{\partial}{\partial \theta} 
Y_{lm}(\theta,0) = {1\over 2} \frac{\partial}{\partial \theta} X_l(p,\theta) \ .
\label{xil1}
\end{eqnarray}
The spherical harmonics have the property that $Y_{l,-m}(\theta,\varphi)=(-1)^m
Y^*_{lm}(\theta,\varphi)$, which implies that $[Y_{l,-m}(\theta,0)]^2=
[Y_{lm}(\theta,0)]^2$. So for $p\ge 1$, we have
\begin{equation}
  X_l(p,\theta)= \sum_{m=1}^l m^p [1+(-1)^p] [Y_{lm}(\theta,0)]^2 \ .
\label{Xl2}
\end{equation}
It follows from Eqs.~(\ref{Xl2}) and (\ref{xil1}) that $X_l(p,\theta)=
\xi_l(p,\theta)=0$ if $p$ is an odd integer.

To evaluate $X_l$, we use the addition formula for spherical harmonics: 
\begin{equation}
  \sum_{m=-l}^l Y^*_{lm}(\theta',\varphi') Y_{lm}(\theta,\varphi) = \frac{2l+1}{4\pi} 
P_l(\nu) \ ,
\label{addYlm}
\end{equation}
where $P_l$ is the Legendre polynomial and 
\begin{equation}
  \nu=\cos\theta \cos\theta' + \sin\theta \sin\theta' \cos(\varphi-\varphi') \ .
\end{equation}
It follows from Eqs.~(\ref{addYlm}), (\ref{Xl1}) and $P_l(1)=1$ that $X_l(0,\theta)=
(2l+1)/(4\pi)$ and $\xi_l(0,\theta)=0$. To compute $X_l$ for $p>1$, we set 
$\theta'=\theta$, $\varphi'=0$ and differentiate 
Eq.~(\ref{addYlm}) $p$ times with respect to $\varphi$. After some 
rearrangement, we obtain
\begin{equation}
  X_l(p,\theta)= \left. (-i)^p \frac{2l+1}{4\pi} \frac{\partial^p}{\partial \varphi^p} 
P_l(\nu) \right|_{\varphi=0} 
\label{Xl3}
\end{equation} 
with $\nu=\cos^2 \theta +\sin^2 \theta \cos \varphi$. For simplicity, we only 
compute $X_l(p,\theta)$ for $p=2$ and $p=4$ here. Generalization to other values 
of $p$ is straightforward. Eq.~(\ref{Xl3}) gives 
\begin{eqnarray}
  X_l(2,\theta) &=& \frac{2l+1}{4\pi}\sin^2 \theta P'_l(1) \\
  X_l(4,\theta) &=& \frac{2l+1}{4\pi}[\sin^2\theta P'_l(1) + 3\sin^4\theta P''_l(1)] \ .
\end{eqnarray}
To compute $P'_l(1)$ and $P''_l(1)$, we use the recurrence relation
\begin{eqnarray}
   P'_{n+1}(x)=xP'_n(x)+(n+1) P_n(x) \ .
\end{eqnarray}
Differentiating the above equation, we have 
\begin{equation}
  P''_{n+1}(x)=xP''_n(x)+(n+2)P'_n(x) \ .
\end{equation}
Evaluating the two equations at $x=1$ and using the fact that $P_n(1)=1$, 
we obtain
\begin{eqnarray}
  P'_{n+1}(1)-P_n'(1) &=& n+1 \label{rec1} \\
  P''_{n+1}(1)-P''_n(1) &=& (n+2) P'_l(1)  \ . \label{rec2}
\end{eqnarray}
Since $P_0(x)=1$, we have $P'_0(1)=P''_0(1)=0$. The difference equations~(\ref{rec1})
and (\ref{rec2}) are then solved by summing over $n$ on both sides from $n=0$ 
to $(l-1)$. The results are 
\begin{eqnarray}
  P'_l(1) &=& \frac{l(l+1)}{2} \\ 
  P''_l(1) &=& \frac{(l-1)l(l+1)(l+2)}{8}  \ .
\end{eqnarray}
Combining our results, we finally have 
\begin{eqnarray}
  X_l(0,\theta) &=& \frac{2l+1}{4\pi} \label{a17}\\
  X_l(2,\theta) &=& \frac{2l+1}{4\pi}\, \frac{l(l+1)}{2}\sin^2\theta \\
  X_l(4,\theta) &=& \frac{2l+1}{4\pi}\left[ \frac{l(l+1)}{2}\sin^2\theta + 
\frac{3}{8}(l-1)l(l+1)(l+2)\sin^4\theta \right] \ .
\end{eqnarray}
The values of $\xi_l(p,\theta)$ can then be computed by Eq.~(\ref{xil1}). The 
results are 
\begin{eqnarray}
  \xi_l(0,\theta) &=& 0 \\
  \xi_l(2,\theta) &=& \frac{2l+1}{4\pi}\, \frac{l(l+1)}{2}\sin\theta \cos\theta \\
  \xi_l(4,\theta) &=& \frac{2l+1}{4\pi} \sin\theta \cos\theta\left[ \frac{l(l+1)}{2}
+ \frac{3}{4}(l-1)l(l+1)(l+2)\sin^2\theta \right] \ .
\label{a22}
\end{eqnarray}

\section{}\label{app-b}
\begin{center}
{\bf EVALUATION OF {\boldmath $f^l_{\varphi}$}}
\end{center}

The $\varphi$-component of the SF $f^l_{\varphi}$ is given by 
\begin{eqnarray}
  f^l_{\varphi} &=& \sum_{m=-l}^l q \partial_{\varphi} \Phi^{lm}  \\ 
 &=& qS(r)\sqrt{M^2-a^2-Q^2} \sum_{m=-l}^l im\frac{\phi^{lm}_1(z) \phi^{lm}_2(z)}
{W_z[\phi^{lm}_1,\phi^{lm}_2]} Y^*_{lm}(\theta,\varphi) Y_{lm}(\theta,\varphi) \ .
\end{eqnarray}
All the quantities are evaluated at the position of the charge, which are assumed 
to be located at $(r,\theta,\varphi)$. Here
\begin{equation}
  W_z[\phi^{lm}_1,\phi^{lm}_2]=\phi^{lm}_1\frac{d\phi^{lm}_2}{dz}-\phi^{lm}_2
\frac{d\phi^{lm}_1}{dz}
\end{equation}
and the factor $\sqrt{M^2-a^2-Q^2}$ comes from the fact that 
\[ 
 \frac{d}{dz}=\sqrt{M^2-a^2-Q^2}\frac{d}{dr} \ .
\]
It is easy to show that $\Phi^{l,-m}=(\Phi^{lm})^*$, so $f^l_{\varphi}$ is 
real. Hence we can write 
\begin{eqnarray}
  f^l_{\varphi} &=& qS(r)\sqrt{M^2-a^2-Q^2} \sum_{m=-l}^l \mbox{Re} \left\{ 
im\frac{\phi^{lm}_1(z) \, \phi^{lm}_2(z)}
{W_z[\phi^{lm}_1,\phi^{lm}_2]} Y^*_{lm}(\theta,\varphi) 
Y_{lm}(\theta,\varphi) \right\}
\cr
 &=& -qS(r)\sqrt{M^2-a^2-Q^2} \sum_{m=-l}^l \mbox{Im} \left\{ 
m\frac{\phi^{lm}_1(z) \, \phi^{lm}_2(z)}
{W_z[\phi^{lm}_1,\phi^{lm}_2]} \right\} 
Y^*_{lm}(\theta,\varphi) Y_{lm}(\theta,\varphi) \ .
\label{flphnew}
\end{eqnarray}

In Sect.~\ref{form}, we find that $\phi^{lm}_1(z) \propto P^{-\mu}_l(z)$ and 
$\phi^{lm}_2(z)\propto Q^{\mu}_l(z)$. However, the function $Q^{-\mu}_l(z)$ is 
related to $Q^{\mu}_l(z)$ by~\cite{am-st}
\begin{equation}
  Q^{-\mu}_l(z)=e^{-2i\mu \pi}\frac{\Gamma(l-\mu+1)}{\Gamma(l+\mu+1)}
Q^{\mu}_l(z) \label{Qlmu-} \ .
\end{equation}
Hence $Q^{-\mu}_l$ and $Q^{\mu}_l$ are linearly dependent, and so we can 
as well choose $\phi^{lm}_2 \propto Q^{-\mu}_l$. In this Appendix, we set 
\begin{eqnarray}
  \phi^{lm}_1(z) &=& P^{-\mu}_l(z) \label{1sol} \\ 
  \phi^{lm}_2(z) &=& Q^{-\mu}_l(z) = \frac{e^{-i\mu \pi}}{(2l+1)!!} 
\, \frac{\Gamma(l-\mu+1)}{z^{l+1}}K^{-\mu}_l(z) \ , \label{2sol}
\end{eqnarray}
where $(2l+1)!!=1\cdot 3 \cdot 5 \cdots (2l+1)$ and 
\begin{equation}
  K^{\mu}_l(z)=\left(1-{1\over z^2}\right)^{\mu/2} { _2F_1}\left(
1+\frac{l+\mu}{2},\frac{l+\mu+1}{2};l+{3\over 2};{1\over z^2}\right) \ .
\label{Klmu}
\end{equation}
We have used Eq.~(\ref{Qlmu}) and the fact that
\begin{equation}
  \Gamma \left(l+{3\over 2}\right) =\frac{(2l+1)!!}{2^{l+1}}\sqrt{\pi} 
\end{equation}
to obtain the second equality in Eq.~(\ref{2sol}). Since $\mu=im\gamma$ 
is purely imaginary, we have $K^{-\mu}_l=(K^{\mu}_l)^*$. It follows from 
Eqs.~(\ref{Qlmu-}) and (\ref{2sol}) that 
\begin{equation}
  K^{-\mu}_l(z)=[K^{\mu}_l(z)]^*=K^{\mu}_l(z) \ ,
\label{Kreal}
\end{equation}
Hence $K^{\mu}_l$ is a real-valued function, although it is not obvious to 
see this from Eq.~(\ref{Klmu}). $Q^{\mu}_l(z)$ is then just a complex factor 
independent of $z$ 
times a real-valued function, so we can choose $\phi^{lm}_2(z)$ to be real. 
In fact, the $\phi^{lm}_2(z)$ in Eq.~(\ref{phi2}) is real, because
it is equal to $K^{\mu}_l(z)/z^{l+1}$. In this Appendix, however, we stick to 
the choice in Eq.~(\ref{2sol}).

The Wronskian is given by~\cite{am-st}
\begin{eqnarray}
  W_z[\phi^{lm}_1,\phi^{lm}_2] &=& \frac{e^{-i\mu\pi}2^{-2\mu}}{1-z^2}\, 
\frac{\Gamma\left( \frac{l-\mu}{2}+1\right) \Gamma\left( \frac{l-\mu+1}{2}\right)}
{\Gamma\left(\frac{l+\mu}{2}+1\right) \Gamma\left(\frac{l+\mu+1}{2}\right)} \cr
 &=& \frac{e^{-i\mu\pi}}{1-z^2}\, \frac{\Gamma(l-\mu+1)}{\Gamma(l+\mu+1)} \ ,
\label{wronz}
\end{eqnarray}
where we have used the identity~\cite{am-st}
\begin{equation}
  \Gamma(2x)=\frac{2^{2x-{1\over 2}}}{\sqrt{2\pi}}\, \Gamma(x) \Gamma\left( x+{1\over 2}
\right) \ .
\end{equation}
Combining Eqs.~(\ref{1sol}), (\ref{2sol}), (\ref{Kreal}) and (\ref{wronz}), we 
obtain 
\begin{equation}
  \frac{\phi^{lm}_1(z)\phi^{lm}_2(z)}{W_z[\phi^{lm}_1,\phi^{lm}_2]} = 
\left[ \frac{1}{z^{l+1}}\, \frac{1-z^2}{(2l+1)!!}\, K^{\mu}_l(z)\right] 
\left[ \Gamma(l+\mu+1) P^{-\mu}_l(z)\right] \ .
\label{ims1}
\end{equation}
Note that the quantity in the first bracket is real, while that in the second 
is complex. To extract the imaginary part, we subtract from Eq.~(\ref{ims1}) 
its complex conjugate and then divide by $2i$:
\begin{eqnarray}
  \mbox{Im}\left[ \frac{\phi^{lm}_1(z)\phi^{lm}_2(z)}{W_z[\phi^{lm}_1,\phi^{lm}_2]} 
\right] &=& \left[ \frac{1}{z^{l+1}}\, \frac{1-z^2}{(2l+1)!!}\, K^{\mu}_l(z)\right]
\left[ \frac{\Gamma(l+\mu+1) P^{-\mu}_l(z)-\Gamma(l-\mu+1) P^{\mu}_l(z)}{2i}\right] 
\cr
 &=& \left[ \frac{1}{z^{l+1}}\, \frac{1-z^2}{(2l+1)!!}\, K^{\mu}_l(z)\right]
\left[-\frac{\Gamma(l+\mu+1)\Gamma(l-\mu+1)}{\pi (2l+1)!!}\, \frac{\sinh (\pi m\gamma)}
{z^{l+1}}\, K^{\mu}_l(z) \right] \ ,
\end{eqnarray}
where we have used the fact that $\mu=im\gamma$ and~\cite{am-st}
\begin{equation}
  P^{\mu}_l(z)=\frac{\Gamma(l+\mu+1)}{\Gamma(l-\mu+1)}\left[ P^{-\mu}_l(z)+
\frac{2}{\pi}e^{i\pi \mu}\sin (\mu \pi)\, Q^{-\mu}_l(z) \right] \ .
\end{equation}
The product $\Gamma(l+\mu+1)\Gamma(l-\mu+1)$ can be evaluated by the identities
$\Gamma(x+1)=x\Gamma(x)$ and~\cite{am-st}
\begin{equation}
  \Gamma(iy)\Gamma(-iy) = \frac{\pi}{y\sinh (\pi y)} 
\end{equation}
for real $y$. The result is 
\begin{equation}
   \mbox{Im}\left[ \frac{\phi^{lm}_1(z)\phi^{lm}_2(z)}{
W_z[\phi^{lm}_1,\phi^{lm}_2]}
\right] = \frac{m\gamma}{z^{2l}}\left(1-{1\over z^2}\right) \left[ 
\prod_{j=1}^l (m^2\gamma^2 +j^2)\right] \left[ \frac{K^{\mu}_l(z)}{(2l+1)!!}
\right]^2 \ .
\end{equation}
Substituting the above formula into Eq.~(\ref{flphnew}), we find   
\begin{equation}
  f^l_{\varphi}=-\frac{q S(r)}{z^{2l}}\gamma \sqrt{M^2-a^2-Q^2} 
\left(1-{1\over z^2}\right) \sum_{m=-l}^l 
m^2 \left[ \prod_{j=1}^l (m^2\gamma^2+j^2)\right]
\left[ \frac{K_l^{\mu}(z)}{(2l+1)!!}\right]^2 Y^*_{lm}(\theta,\varphi) 
Y_{lm}(\theta,\varphi) \ ,
\end{equation}
and substituting for the value of $\gamma$ we finally obtain 
\begin{equation}
  f^l_{\varphi}=-\frac{q S(r)}{z^{2l}} a 
\left(1-{1\over z^2}\right) \sum_{m=-l}^l
m^2 \left[ \prod_{j=1}^l (m^2\gamma^2+j^2)\right]
\left[ \frac{K_l^{\mu}(z)}{(2l+1)!!}\right]^2 Y^*_{lm}(\theta,\varphi)
Y_{lm}(\theta,\varphi) \ .
\end{equation}

\section{}\label{ap-num}
\begin{center}
{\bf NUMERICAL EVALUATION OF THE SELF FORCE}
\end{center}

A very effective way to evaluate the associated Legendre functions
$P_{l}^{-\mu}(z)$ and $Q_{l}^{\mu}(z)$ numerically is by their
representations using hypergeometric functions. Specifically, we rewrite
the associated Legendre functions as
\begin{equation}
P_{l}^{-\mu}(z)=\frac{1}{\Gamma(1+\mu)}\left(\frac{z+1}{z-1}\right)
^{-\mu/2}{{_2}F_{1}}\left(-l,l+1;1+\mu;\frac{1-z}{2}\right)
\label{Phg}
\end{equation}
and
\begin{equation}
Q_{l}^{\mu}(z)=e^{i\mu\pi}
\frac{\sqrt{\pi}}{2^{l+1}}\frac{\Gamma(1+l+\mu)}
{\Gamma(l+3/2)}\frac{(z^2-1)^{\mu/2}}{z^{1+l+\mu}}\, 
{{_2}F_{1}}\left(1+\frac{l+\mu}{2},\frac{1+l+\mu}{2};
l+\frac{3}{2};\frac{1}{z^2}\right)\, .
\label{Qhg}
\end{equation}   
In all the expressions we need to evaluate, we only have products of 
$P_{l}^{-\mu}(z)$ and $Q_{l}^{\mu}(z)$ divided by their Wronskian
determinant. Hence, we do not need to evaluate the constant factors in
Eqs.\ (\ref{Phg}) and (\ref{Qhg}). The derivatives of $P_{l}^{-\mu}(z)$
and $Q_{l}^{\mu}(z)$, which are needed both for the Wronskian
determinant and for the gradient of the field in the SF computation, can
be computed using the relation
\begin{equation}
\frac{d}{\,dz}\, {{_2}F_{1}}\left(a,b;c;z\right)=\frac{ab}{c}\, 
{{_2}F_{1}}\left(a+1,b+1;c+1;z\right)\, .
\label{HGder}
\end{equation}
It is convenient to evaluate the hypergeometric functions in Eqs.\ 
(\ref{Phg}-\ref{HGder}) using their Gauss series representation, i.e., 
\begin{equation}
{{_2}F_{1}}\left(a,b;c;z\right)=
\sum_{n=0}^{\infty}\frac{(a)_n(b)_n}{n!(c)_n}z^n \ \ \
|z|<1 \, ,
\label{c4}
\end{equation}
where $(a)_{n}$ is Pochhammer's symbol. This is a useful
numerical approach because of the following. For the
evaluation of $P_{l}^{-\mu}(z)$ we need to compute a hypergeometric
function for which the Gauss series is reduced to a polynomial of degree
$l$ in $(1-z)/2$, thanks to the first variable of the hypergeometric
function being a negative integer. Hence, we only need to sum over a
finite number of terms, and there is no truncation error involved. 
We are also unrestricted by the radius of convergence of (\ref{c4}). 
This fortunate circumstance does not change for the calculation of the
derivative. For the computation of $Q_{l}^{\mu}(z)$ we benefit from the
argument being positive and smaller than unity. Consequently, in the Gauss
series we have a sum over terms which are a 
$({\rm combinatorial\; factor})\times z^{-2n}$, where $z>1$ and $n$ is a
positive integer. This guarantees that the series converges fast, such
that in practice we do not need to sum over too many terms to ensure a
given accuracy. In practice we evaluate all hypergeometric function to
accuracy of 1 part in $10^{16}$. Our computation does not become more
complicated because of $\mu$ being imaginary. 

An alternative approach for computing $Q_l^{\mu}(z)$ is to use its  
integral representation, given by
\begin{equation}
Q_l^{\mu}(z)=e^{i\mu\pi}\frac{\Gamma(l+\mu+1)}{2^{l+1}\Gamma(l+1)}
(z^2-1)^{\mu/2}\int_{-1}^{1}\,dt \frac{(1-t^2)^l}{(z-t)^{l+\mu+1}}\, ,
\end{equation}
which is efficient numerically, because the integration interval is 
compact, and the integrand has pathologies neither inside the interval nor 
at its boundaries. 

The spherical harmonics can be computed either from the associated
Legendre functions computed using the hypergeometric representation above
(which is not very convenient because of the large arguments of the gamma
functions, which are needed to be computed now), or, alternatively, from
the recurrence relations for the associated Legendre functions as given in
\cite{num-recipes}. This latter approach is stable and accurate also for
the very large values of $l,m$ we need ($\approx 1000$).

\end{appendix}

\end{document}